\pdfoutput=1

\documentclass[sigconf]{acmart}

\usepackage{graphicx}
\usepackage{balance}
\usepackage{comment}
\usepackage{amsmath}
\usepackage{amssymb}
\usepackage{color}

\usepackage{subfigure}
\usepackage{algorithm}
\usepackage{algorithmic}
\usepackage{refcount}
\usepackage{multirow}

\newcommand{\sect}{Section~}

\graphicspath{{./figs/}}

\copyrightyear{2017}
\acmYear{2017}
\setcopyright{acmcopyright}
\acmConference{SenSys '17}{November 6--8, 2017}{Delft,
Netherlands}\acmPrice{15.00}\acmDOI{10.1145/3131672.3131681}
\acmISBN{978-1-4503-5459-2/17/11}

\begin{document}

\title[Clock Synchronization for Wearables Using SEP Induced by Powerline Radiation]{Application-Layer Clock Synchronization for Wearables Using Skin Electric Potentials Induced by Powerline Radiation}

\author{Zhenyu Yan}
\affiliation{%
  \institution{School of Computer Science and Engineering\\Nanyang Technological University}
  \streetaddress{50 Nanyang Avenue}
  \country{Singapore}
  \postcode{639798}
}
\email{zyan006@ntu.edu.sg}

\author{Yang Li}
\affiliation{%
  \institution{Advanced Digital Sciences Center\\Illinois at Singapore}
  \streetaddress{1 Fusionopolis Way, \#08-10 Connexis North Tower}
  \country{Singapore} 
  \postcode{138632}
}
\email{yang.li@adsc.com.sg}

\author{Rui Tan}
\authornote{Corresponding author.}
\affiliation{%
  \institution{School of Computer Science and Engineering\\Nanyang Technological University}
  \streetaddress{50 Nanyang Avenue}
  \country{Singapore}
  \postcode{639798}
}
\email{tanrui@ntu.edu.sg}

\author{Jun Huang}
\affiliation{
  \institution{Center for Energy Efficient Computing and Applications\\Peking University}
  \streetaddress{5 Yiheyuan Road}
  \city{Haidian}
  \state{Beijing}
  \country{China}
  \postcode{100871}
}
\email{jun.huang@pku.edu.cn}

\begin{abstract}
Design of clock synchronization for networked nodes faces a fundamental trade-off between synchronization accuracy and universality for heterogeneous platforms, because a high synchronization accuracy generally requires platform-dependent hardware-level network packet timestamping. This paper presents {\em TouchSync}, a new indoor clock synchronization approach for wearables that achieves millisecond accuracy while preserving universality in that it uses standard system calls only, such as reading system clock, sampling sensors, and sending/receiving network messages. The design of TouchSync is driven by a key finding from our extensive measurements that the skin electric potentials (SEPs) induced by powerline radiation are salient, periodic, and synchronous on a same wearer and even across different wearers. TouchSync integrates the SEP signal into the universal principle of Network Time Protocol and solves an integer ambiguity problem by fusing the ambiguous results in multiple synchronization rounds to conclude an accurate clock offset between two synchronizing wearables. With our shared code, TouchSync can be readily integrated into any wearable applications. Extensive evaluation based on our Arduino and TinyOS implementations shows that TouchSync's synchronization errors are below 3 and 7 milliseconds on the same wearer and between two wearers 10 kilometers apart, respectively.
\end{abstract}



\maketitle

\section{Introduction}
\label{sec:intro}

The annual worldwide shipments of consumer wearables (e.g., smart watches, wristbands, eyewears, clothing, etc) have grown by 29\% in 2016 \cite{idc-wearable}. This rapid growth is expected to continue, projecting to 213 million units shipped in 2020 \cite{idc-wearable}. Along with the proliferation of consumer wearables, specialized domains such as clinical/home healthcare \cite{chan2012smart} and exercise/sport analysis \cite{mokaya2016burnout} are also increasingly adopting smart wearable apparatuses. In the body-area networks formed by these wearables, a variety of system functions and applications depend on tight clock synchronization among the nodes. For instance, two earbuds of a wireless headphone
need to be synchronized mutually and/or with a master device (e.g., a smartphone) to control the playback positions in their buffers to deliver audio synchronously
\cite{dinescu2015synchronizing}. Motion analysis \cite{lorincz2009mercury} and muscle activity monitoring \cite{mokaya2015myovibe,mokaya2016burnout} require sensory data from multiple tightly synchronized nodes.

While current wearable systems adopt customized, proprietary clock synchronization approaches \cite{apple-sync}, we envisage a wide spectrum of interoperable wearables that can synchronize with each other to enable more novel applications. For instance, in body sensor assisted multi-user gaming that may need to decide which participant performs an action or gesture first, tight clock synchronization among the body sensors and/or the handheld game consoles is needed. In the envisaged scheme, an application developer can readily synchronize any two communicating wearables using high-level and standard system calls provided by their operating systems (OSes), such as reading system clock, transmitting and receiving network messages.
However, the design of clock synchronization approaches faces a fundamental trade-off between the synchronization accuracy and the universality for heterogeneous platforms. This is because a high synchronization accuracy generally requires low-level timestamping for the synchronization packets, which may be unavailable on the hardware platforms or inaccessible to the application developer.

We illustrate this accuracy-universality trade-off using the Network Time Protocol (NTP) \cite{ntp} and the Precision Time Protocol (PTP) \cite{4579760}. NTP synchronizes a slave node and a master node by recording their clock values when a UDP synchronization packet is passed to and received from the sender's and receiver's OSes, respectively. Thus, NTP is universal in that it can be applied to any host that speaks UDP. However, as the application-layer timestamping cannot capture the details of the nondeterministic OS and network delays, NTP may yield significant synchronization errors up to hundreds of milliseconds (ms) in a highly asymmetric network. To solve this issue, PTP uses hardware-level timestamping provided by PTP-compatible Ethernet cards at the end hosts and all the switches on the network path to achieve microsecond ($\mu$s) accuracy. However, the need of the special hardware inevitably negates its universality and restricts PTP's adoption to time-critical local-area networks only, e.g., those found in industrial systems.

In wireless networks, due to the more uncertain communication delays caused by media access control (MAC), NTP performs worse. Thus, similar to PTP, most existing clock synchronization approaches for wireless sensor networks (WSNs) (e.g., RBS \cite{elson2002fine}, TPSN \cite{ganeriwal2003timing}, and FTSP \cite{maroti2004flooding}) have resorted to MAC-layer timestamping provided by the nodes' radio chips to pursue synchronization accuracy.
While FTSP has become the {\em de facto} standard in TinyOS-based WSNs, the need of the MAC-layer timestamping presents a significant barrier for its wide adoption to the broader Internet of Things (IoT) domain, where the IoT platforms use diverse radios and OSes, and in general they do not provide an interface for the MAC-layer timestamping.

In this paper, we aim at developing a new clock synchronization approach for wearables that establishes a desirable accuracy-universality trade-off point between the two extremes represented by NTP and PTP to well address the momentum of IoT platform heterogenization. In particular, we will stem from the sensor nature of wearables to explore ambient signals that can assist clock synchronization.
Recent studies exploited external periodic signals such as powerline radiation \cite{Rowe2009lowpower} and Radio Data System (RDS) \cite{li2011exploiting} to calibrate the clocks of WSN nodes. However, these approaches need non-trivial extra hardware to capture the external signals. Moreover, they focus on {\em clock calibration} that ensures different clocks advance at the same speed, rather than synchronizing the clocks to have the same value. But they inspire us to inquire (i) the existence of a periodic and synchronous signal that can be sensed by different wearables without adding non-trivial hardware to preserve universality and (ii) how to exploit the signal to synchronize the wearables without using hardware-level packet timestamping.

For the first inquiry, we conduct extensive measurements to explore such signals. Our measurements based on Adafruit's Flora \cite{flora}, an Arduino-based wearable platform, show that by simply sampling an onboard analog-to-digital converter (ADC), a Flora can capture powerline electromagnetic radiation that oscillates at the power grid frequency (e.g., $50\,\text{Hz}$). When the Flora's ADC has a physical contact with the wearer's skin, the sampled skin electric potential (SEP) is a significantly amplified version of the powerline radiation, because the human body acts as an effective antenna. Although the SEP's amplitude is dynamic due to the human body movements, its frequency is highly stable. Moreover, the SEPs on the same wearer and even different wearers in a same indoor environment exhibit desirable synchronism. The time displacement between the SEPs at different positions of a still wearer is generally less than $1\,\text{ms}$. These results suggest that SEP is a promising basis for synchronizing the wearables.

For the second inquiry, we integrate the periodic and synchronous SEP signal into the universal principle of NTP to deal with the major source of NTP's error, i.e., asymmetric communication delays. In the original NTP, the problem of estimating the offset between the slave's and master's clocks is a {\em real-domain} underdetermined problem that has infinitely many solutions. NTP chooses a solution by assuming symmetric communication delays, which does not hold in many scenarios, however. Assisted with the periodic SEP, the problem reduces to an {\em integer-domain} underdetermined problem that has a finite number of solutions. However, it is challenging to infer which solution is correct. In this paper, we show that, if the time displacement between the SEP signals captured by two synchronizing wearables is shorter than half of the power grid voltage cycle, the integer ambiguity can be resolved by jointly considering multiple synchronization rounds with dynamic and asymmetric communication delays.
Thus, the clock offset between a pair of wearables can be estimated with ms accuracy due to the ms synchronism between their SEP signals.

Based on the above two key results, we design a novel clock synchronization approach for wearables, which we call {\em TouchSync}. It runs at the application layer in that the needed SEP sampling, the network message exchange and timestamping can be implemented using standard wearable OS calls. Thus, by introducing a rather simple skin contact, we can readily achieve ms synchronization accuracy across heterogeneous wearable platforms, without resorting to the hardware-level packet timestamping that is extremely difficult to be standardized. To simplify the adoption of TouchSync by application developers, we design and release a \texttt{touchsync.h} header file \cite{touchsync-h} that implements TouchSync's signal processing algorithms and the integer ambiguity solver. With this header, the implementations of TouchSync in Arduino and TinyOS need about 50 and 150 lines of code, respectively, which manage sensor sampling, synchronization message exchange and application-layer timestamping only. All these tasks are basics for Arduino and TinyOS developers. We also conduct extensive experiments in various indoor environments to show the pervasive availability of the SEP signals. On the same wearer, TouchSync mostly achieves sub-ms accuracy and the largest error is $2.9\,\text{ms}$. We also conduct a TouchSync-over-Internet proof-of-concept experiment that yields errors below $7\,\text{ms}$ between two wearers $10\,\text{km}$ apart.

The ADC-skin contact needed by TouchSync can be easily integrated into the wearable designs with near-zero cost. Although our experiments show that, in the absence of the contact, TouchSync can still work with graceful performance degradation, SEP is a new and valuable sensing modality for integration consideration, since it provides accurate timing and is indicative of other information about the wearer (e.g., body orientation, movements, and indoor location) as suggested by our measurements in this paper.

The rest of this paper is organized as follows. \sect\ref{sec:related} reviews related work. \sect\ref{sec:background} introduces background and our objective. \sect\ref{sec:measurement} presents a measurement study. \sect\ref{sec:approach}, \sect\ref{sec:implementation}, and \sect\ref{sec:evaluation} designs, implements, and evaluates TouchSync, respectively. \sect\ref{sec:conclude} concludes.

\section{Related Work}
\label{sec:related}

Highly stable time sources are ill-suited for wearables. Chip-scale atomic clocks are too expensive (\$1,500 per unit \cite{chipatomic}). GPS receivers are power-consuming and do not work in indoor environments. Recent studies exploited external signals available in indoor environments to synchronize or calibrate the clocks of distributed nodes. In \cite{Chen2011ultralow}, an AM radio receiver is designed to decode the global time information broadcast by timekeeping radio stations. In \cite{li2011exploiting}, a mote peripheral is designed to capture the periodic RDS signals of FM radios for clock calibration. In \cite{Rowe2009lowpower}, a mote peripheral called syntonistor can receive the periodic electromagnetic radiation from powerlines to calibrate wireless sensors' clocks, where some clock synchronization approach is still needed for the initial synchronization. In \cite{sreejaya16,dima17}, voltage sensors plugged in to wall power outlets are used to secure clock synchronization against malicious network delays. In particular, the voltage cycle length fluctuations are exploited to implement a data-based clock synchronization approach \cite{sreejaya16}. In \cite{yang17ipsn}, such fluctuations extracted from the powerline radiation are used as {\em natural timestamps}. However, the error of the natural timestamps can be up to hundreds of ms. Moreover, the clock synchronization based on the natural timestamps needs to transmit a considerable amount of cycle length data and a compute-intensive matching process to decode the fluctuations to time information \cite{yang17ipsn}. Thus, the natural timestamping approach is ill-suited for tight clock synchronization among resource-constrained wearables.
In \cite{lazik2015ultrasonic}, a smartphone captures ultrasonic beacons from pre-deployed synchronized beacon nodes to synchronize its own clock. All the above approaches \cite{Chen2011ultralow,li2011exploiting,Rowe2009lowpower,sreejaya16,dima17,yang17ipsn,lazik2015ultrasonic} need non-trivial customized hardware and infrastructures, which reduce their universality.

Two recent studies leverage built-in sensing modalities to capture external periodic signals for clock calibration. In \cite{hao2011wizsync}, a 802.15.4 radio is used to capture the Wi-Fi beacons to calibrate motes' clocks. Although this approach does not require a peripheral, it uses the received signal strength indication (RSSI) register of the CC2420 radio chip, which makes it hardware specific and nonuniversal for IoT platforms that use diverse radios.
In \cite{li2012flight}, motes use light sensors to capture the fluorescent light that flickers at a frequency twice of the power grid frequency to calibrate their clocks. Although light sensors are widely available, the required fluorescent lighting may not be available in natural lighting environment. In contrast, the powerline radiation that induces the SEP signal used by our approach pervades civil infrastructures.

The studies \cite{li2011exploiting,Rowe2009lowpower,hao2011wizsync,li2012flight} mentioned above, including the two \cite{Rowe2009lowpower,li2012flight} that are power grid related, focus on {\em clock calibration} that involves no message exchanges among nodes. Though continuous clock calibration maintains the nodes synchronized once they are initially synchronized, the initial synchronization and the resynchronizations needed upon clock calibration faults are not addressed in these studies. Thus, these studies and ours are complementary, in that the principle of TouchSync can be used for the initial synchronization of the systems adopting these clock calibration approaches \cite{li2011exploiting,Rowe2009lowpower,hao2011wizsync,li2012flight}.

\section{Background and Objective}
\label{sec:background}

\subsection{NTP Principle and Packet Timestamping}
\label{subsec:ntp}

\begin{figure}
  \subfigure[NTP principle.]
  {
    \includegraphics{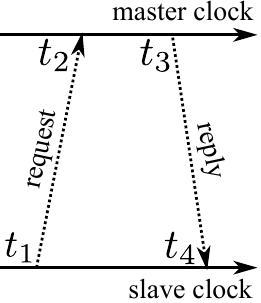}
    \label{fig:ntp}
  }
  \hspace{.5em}
  \subfigure[Packet timestamping for synchronization.]
  {
    \includegraphics{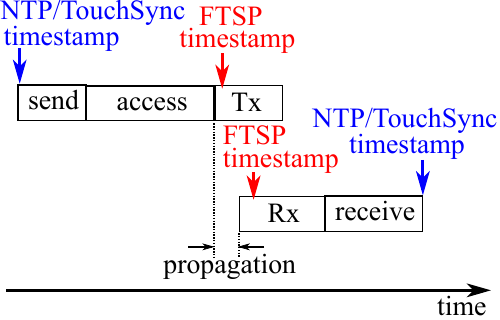}
    \label{fig:timestamping}
  }
  \caption{NTP principle and packet timestamping.}
\end{figure}

Many clock synchronization approaches adopt the principle of NTP, which is illustrated in Fig.~\ref{fig:ntp}. A {\em synchronization session} consists of the transmissions of a request packet and a reply packet. The $t_1$ and $t_4$ are the slave's clock values when the request and reply packets are transmitted and received by the slave node, respectively. The $t_2$ and $t_3$ are the master's clock values when the request and reply packets are received and transmitted by the master node, respectively. Thus, the round-trip time (RTT) is $\mathrm{RTT} = (t_4 - t_1) - (t_3 - t_2)$. Based on a {\em symmetric link assumption} that assumes identical times for transmitting the two packets, the offset between the slave's and the master's clocks, denoted by $\delta_{NTP}$, is estimated as $\delta_{NTP} = t_4 - \left(t_3 + \frac{\mathrm{RTT}}{2}\right)$. Then, this offset is used to adjust the slave's clock to achieve clock synchronization. Under the above principle, non-identical times for transmitting the two packets will result in an error in estimating the clock offset. The estimation error is half of the difference between the times for transmitting the two packets.


We use Fig.~\ref{fig:timestamping} and the terminology in \cite{maroti2004flooding} to explain how the timestamps (e.g., $t_3$ and $t_4$) are obtained in NTP and existing WSN clock synchronization approaches. The {\em send time} and the {\em receive time} are the times used by the OS to pass a packet between the synchronization program and the MAC layer at the sender and receiver, respectively. They depend on OS overhead. The {\em access time} is the time for the sender's MAC layer to wait for a prescribed time slot in time-division multiple access (TDMA) or a clear channel in carrier-sense multiple access with collision avoidance (CSMA/CA). It often bears the highest uncertainty and can be up to $500\,\text{ms}$ \cite{maroti2004flooding}. The transmission (Tx) and reception (Rx) times are the physical layer processing delays at the sender and receiver, respectively. The {\em propagation time} equals the distance between the two nodes divided by the speed of light, which is generally below $1\,\mu\text{s}$.

As illustrated in Fig.~\ref{fig:timestamping}, NTP timestamps the packet when the packet is passed to or received from the OS. Thus, the packet transmission time used by NTP is subjected to the uncertain OS overhead and MAC. Therefore, as measured in \sect\ref{subsec:ntp-perf}, NTP over a Bluetooth connection can yield nearly $200\,\text{ms}$ clock offset estimation errors. To remove these uncertainties, FTSP uses MAC-layer access to obtain the times when the beginning of the packet is transmitted/received by the radio chip. As the propagation time is generally below $1\,\mu\text{s}$, FTSP simply estimates the clock offset as the difference between the two hardware-level timestamps. Thus, the two-way scheme in Fig.~\ref{fig:ntp} becomes non-essential for FTSP.

\subsection{Objective}

NTP, though universal, gives unacceptably low accuracy. On the other extreme, existing WSN clock synchronization approaches, though achieving $\mu$s accuracy, may not be universally applicable to the diversified IoT platforms with different radio chips and OSes. In this paper, by introducing the readily available SEP signal, we aim at developing a new clock synchronization approach for wearables that (i) uses application-layer timestamping as NTP does to preserve universality and (ii) achieves ms accuracy that meets the requirements of a range of applications. For instance, music streaming requires a synchronism below $30\,\text{ms}$ between two wireless earbuds \cite{dinescu2015synchronizing}. For seismic sensing based motion analysis \cite{lorincz2009mercury} and muscle activity monitoring \cite{mokaya2016burnout} that generally adopt sampling rates up to $500\,\text{Hz}$, the ms synchronization accuracy can enable us to discriminate any readings sampled by different sensors at different time instants. To this end, we need to understand the properties of SEP, which is the subject of \sect\ref{sec:measurement}.

\section{Measurement Study}
\label{sec:measurement}

In this section, we conduct measurements to gain insights for guiding the design of TouchSync.

\subsection{Measurement Setup}
\label{subsec:hardware}

\begin{figure}
  \begin{minipage}[t]{.18\textwidth}
    \includegraphics[width=\textwidth]{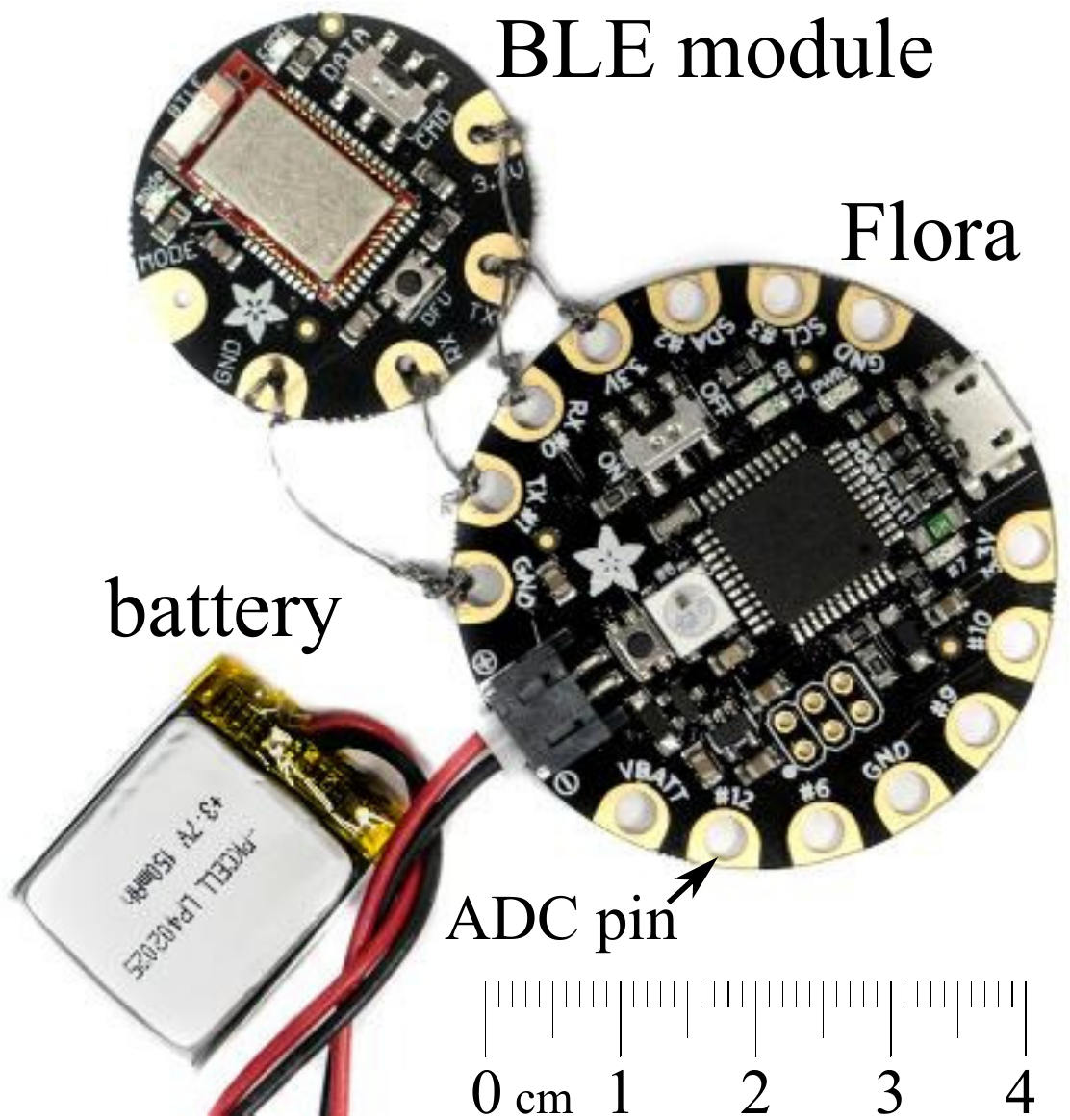}
    \caption{Flora.}
    \label{fig:flora}
  \end{minipage}
  \hspace{2em}
  \begin{minipage}[t]{.21\textwidth}
    \includegraphics[width=\textwidth]{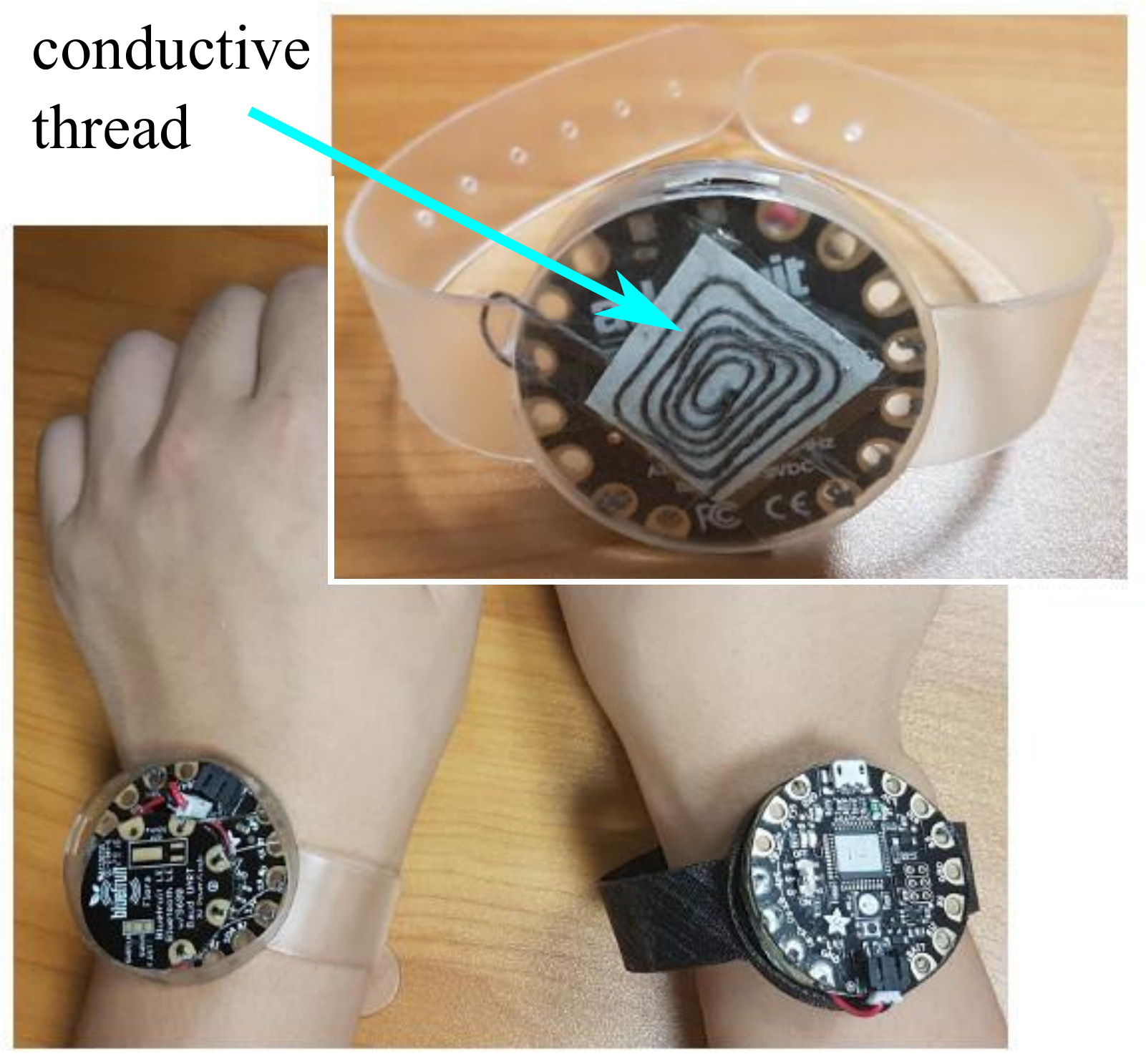}
    \caption{Prototypes supporting TouchSync.}
    \label{fig:prototype}
  \end{minipage}
\end{figure}

Our measurement study uses two Adafruit Flora nodes \cite{flora} and a Raspberry Pi (RPi) 3 Model B single-board computer \cite{rpi3}. The Flora is an Arduino-based wearable platform that can be programmed using the Arduino IDE. Each Flora node, as shown in Fig.~\ref{fig:flora}, consists of a main board with an ATmega32u4 MCU ($8\,\text{MHz}$, $2.5\,\text{KB}$ RAM), a Bluetooth Low Energy (BLE) 4.1 module, and a 150mAh lithium-ion polymer battery. The RPi has a built-in BLE 4.1 module and runs Ubuntu MATE 16.04 with BlueZ 5.37 as the BLE driver. We use Adafruit's nRF51 Arduino library \cite{adafruit-nrf51} and BluefruitLE Python library \cite{adafruit-bluefruitle} on the Floras and the RPi, respectively, to send and receive data over BLE in the UART mode through the \texttt{write()} and \texttt{read()} functions. The Floras and RPi operate as BLE peripheral (slave) and central (master), respectively. To obtain the ground truth clocks, in each experiment, we synchronize the Floras with the RPi as follows. At the beginning of the experiment, we wire a general-purpose input/output (GPIO) pin of the RPi with a digital input pin of each Flora. Then, the RPi issues a rising edge through the GPIO pin and records its clock value $t_{master}$. Upon detecting the rising edge, a Flora records its clock value $t_{slave}$ and sends it to the RPi. The RPi computes the ground-truth offset between the Flora's and the RPi's clocks as $\delta_{GT}=t_{slave}-t_{master}$. Then, we remove the wiring and conduct experiments.

\subsection{Performance of NTP over BLE Connection}
\label{subsec:ntp-perf}

\begin{figure}
\subfigure[]
{
  \includegraphics{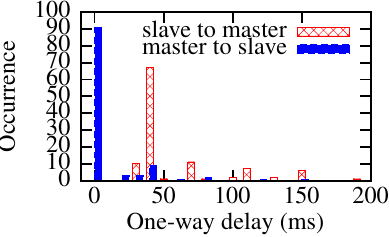}
  \label{fig:one-way-delays}
}
\subfigure[]
{
  \includegraphics{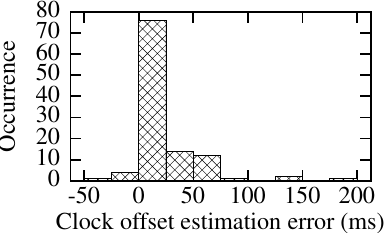}
  \label{fig:ntp-error}
}
\caption{Performance of NTP over a BLE connection.}
\end{figure}

As our objective is to devise a new clock synchronization approach that uses application-layer timestamping as NTP does, this section measures the performance of NTP to provide a baseline. We implement the NTP principle described in \sect\ref{subsec:ntp} on the Flora setup. Fig.~\ref{fig:one-way-delays} shows the distributions of the one-way application-layer communication delays over 110 NTP sessions. The slave-to-master delays are mostly within $[40, 50]\,\text{ms}$, with a median of $42\,\text{ms}$ and a maximum of $376\,\text{ms}$ (not shown in the figure). As specified by the BLE standard, the master device pulls data from a slave periodically. The period, called {\em connection interval}, is determined by the master. The slave needs to wait for a pull request to transmit a packet to the master. In BlueZ, the connection interval is set to $67.5\,\text{ms}$ by default. As the arrival time of a packet from the slave's OS is uniformly distributed over the connection interval, the expected access time is $67.5 / 2 = 33.75\,\text{ms}$. This is consistent with our measured median delay of $42\,\text{ms}$, which is about $8\,\text{ms}$ longer because of other delays (e.g., send and receive times). The exceptionally long delays (e.g., $376\,\text{ms}$) observed in our measurements could be caused by transient wireless interference and OS delays. For the master-to-slave link, the delays are mostly within $[0, 10]\,\text{ms}$, with a median of $8\,\text{ms}$ and a maximum of $153\,\text{ms}$. A BLE slave can skip a number of pull requests, which is specified by the {\em slave latency} parameter, and sleep to save energy. Under BlueZ's default setting of zero for slave latency, the slave keeps awake and listening, yielding short master-to-slave delays.

The asymmetric slave-to-master and master-to-slave delays will cause significant errors in the NTP's clock offset estimation. At the end of each synchronization session, the RPi computes this error as $\delta_{NTP} - \delta_{GT}$, where $\delta_{NTP}$ and $\delta_{GT}$ are NTP's estimate and the ground-truth offset, respectively. Fig.~\ref{fig:ntp-error} shows the distribution of the errors. We observe that 28\% of the errors are larger than $25\,\text{ms}$. The largest error in the 110 NTP sessions is $183\,\text{ms}$. Such an error profile does not well meet the ms accuracy requirements of many applications \cite{dinescu2015synchronizing,lorincz2009mercury,mokaya2016burnout}. Though it is possible to calibrate the average error to zero by using prior information (e.g., the settings of the connection interval and slave latency), the calibration is tedious, nonuniversal, and incapable of reducing noise variance.

\subsection{Skin Electric Potential (SEP)}
\label{sec:body-antenna}

In this set of measurements, we explore i) whether a human body is an effective antenna for receiving the powerline radiation and ii) whether the SEP signals induced by the radiation on the same wearer or different wearers are synchronous. The Flora's MCU has a 10-bit ADC that supports a sampling rate of up to $15\,\text{kHz}$. To facilitate experiments, we have made two Flora-based prototypes as shown in Fig.~\ref{fig:prototype}. We place the Flora into a 3D-printed insulating wristband and use a stainless thin conductive thread to create a connection between Flora's ADC pin and the wearer's skin.
The Flora samples the ADC at $333\,\text{Hz}$ continuously for two minutes and streams the timestamped raw data to the RPi for offline analysis. The sampling rate of $333\,\text{Hz}$ is sufficient to capture the powerline radiation or SEP with a frequency of $50\,\text{Hz}$ in our region. All samples are normalized using the reference voltage of the ADC. As only the ADC pin is connected to the researcher, the grounding of the Flora may affect the sampling result. To understand the impact of grounding, we conduct two sets of comparative experiments, where the two Floras have shared and independent grounds, respectively. In each experiment set, there are two scenarios: still and moving. For the moving scenario, the researcher keeps changing the body orientation, movement, and location. The experiments are conducted in a computer science laboratory with various appliances such as lights, computers, and printers. 

\subsubsection{Shared ground}
\label{subsubsec:shared-ground}

We wire the ground pins of the two Floras, such that they have a shared ground. We conduct three tests.

\begin{figure}
  \subfigure[Shared ground]
  {
    \includegraphics{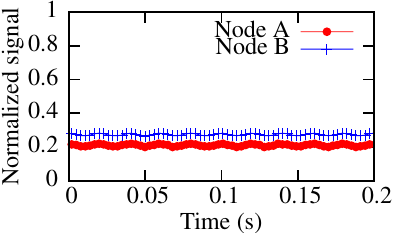}
    \label{fig:raw-nobody-shared}
  }
  \subfigure[Independent grounds]
  {
    \includegraphics{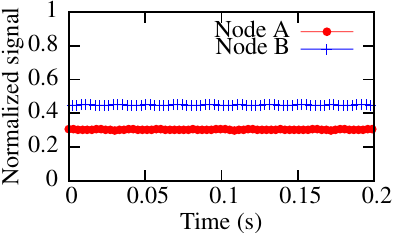}
    \label{fig:raw-nobody-independent}
  }
  \caption{No human body contact.}
\end{figure}

First, Fig.~\ref{fig:raw-nobody-shared} shows the signals captured by the two Floras when they have no physical contact with any human body. The signals have small fluctuations with a normalized peak-to-peak amplitude of 0.024. The signals fluctuate at a frequency of $50\,\text{Hz}$. This suggests that the Floras can pick up the powerline radiation. However, the signals are weak.

\begin{figure}
  \subfigure[Still body]
  {
    \includegraphics{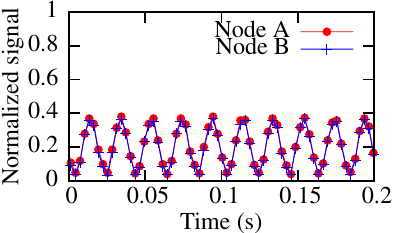}
    \label{fig:raw-samebody-static}
  }
  \subfigure[Moving body]
  {
    \includegraphics{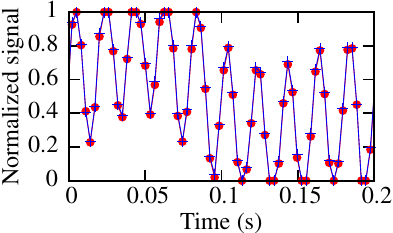}
    \label{fig:raw-samebody-shaking}
  }
  \caption{SEPs on a same wearer (shared ground).}
  \label{fig:raw-samebody}
\end{figure}

Second, a researcher touches the ADC pins of the two Floras with his two hands, respectively. Figs.~\ref{fig:raw-samebody-static} and \ref{fig:raw-samebody-shaking} show the signals captured by the two nodes during the same time duration, when the researcher stands still and walks, respectively. Under the two scenarios (still and moving), the two nodes yield salient and almost identical signals. The peak-to-peak amplitudes in the two figures are around 0.4 and 0.8, which are 17 and 33 times larger than that of the signal shown in Fig.~\ref{fig:raw-nobody-shared}. This suggests that the human body can effectively receive the powerline radiation.

\begin{figure}
  \subfigure[Still bodies]
  {
    \includegraphics{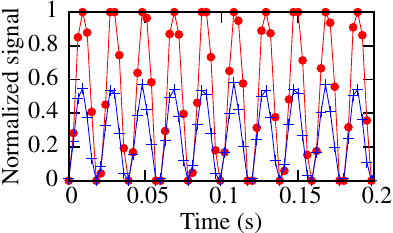}
    \label{fig:raw-diffbody-static}
  }
  \subfigure[Moving bodies]
  {
    \includegraphics{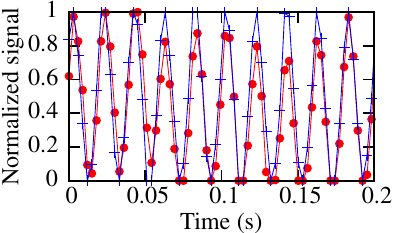}
    \label{fig:raw-diffbody-shaking}
  }
  \caption{SEPs on different wearers (shared ground).}
  \label{fig:raw-diffbody}
\end{figure}

Third, two researchers touch the ADC pins of the two Floras separately. Figs.~\ref{fig:raw-diffbody-static} and \ref{fig:raw-diffbody-shaking} show the signals captured by the two nodes when the two researchers stand still and walk, respectively. The two nodes yield salient signals with different amplitudes. We note that several factors may affect the reception of powerline radiation, e.g., human body size, position and facing of the body in the electromagnetic field generated by the powerlines.

\begin{figure}
  \subfigure[Shared ground]
  {
    \includegraphics{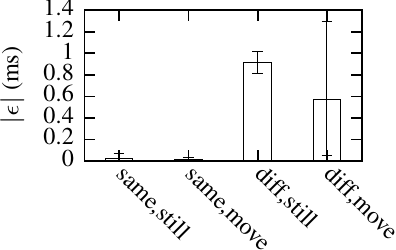}
    \label{fig:delay-shared-ground}
  }
  \subfigure[Independent grounds]
  {
    \includegraphics{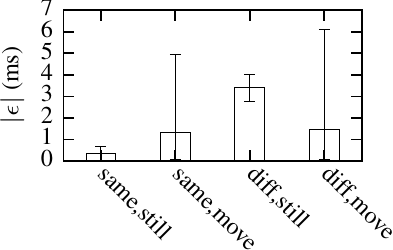}
    \label{fig:delay-independent-ground}
  }
  \caption{Absolute time displacement $|\epsilon|$ between the EMR signals captured by the two Floras in various scenarios. Error bar represents $(5\%,95\%)$ confidence interval. Each error bar is obtained from one minute of data.}
  \label{fig:epsilon}
\end{figure}

We evaluate the synchronism between the signals captured by the two Floras shown in Figs.~\ref{fig:raw-samebody} and \ref{fig:raw-diffbody}. We condition the signals by first applying a band-pass filter (BPF) to remove the direct current (DC) component that may fluctuate as seen in Fig.~\ref{fig:raw-samebody-shaking} and then detect the zero crossings (ZCs) of the filtered signals. More details of the BPF and ZC detection will be presented in \sect\ref{subsec:conditioning}. We use the {\em time displacement} between the two signals' ZCs as the metric to evaluate their synchronism. Specifically, the time displacement, denoted by $\epsilon$, is given by $\epsilon = t_{A}^{ZC} - t_{B}^{ZC}$, where $t_{A}^{ZC}$ and $t_{B}^{ZC}$ represent the ground-truth times of Node $A$'s ZC and the corresponding ZC at Node $B$, respectively. Fig.~\ref{fig:delay-shared-ground} shows the error bars for $|\epsilon|$, which correspond to the scenarios in Figs.~\ref{fig:raw-samebody-static}, \ref{fig:raw-samebody-shaking}, \ref{fig:raw-diffbody-static}, and \ref{fig:raw-diffbody-shaking}, respectively. In Fig.~\ref{fig:delay-shared-ground}, ``same'' and ``diff'' mean the same wearer and different wearers, respectively; ``still'' and ``move'' mean standing still and walking, respectively. On the same wearer, the SEPs captured by the two Floras are highly synchronous, with an average $|\epsilon|$ of $0.9\,\mu\text{s}$. On different wearers, the $|\epsilon|$ increases to about $1\,\text{ms}$. When the two wearers move, the average $|\epsilon|$ is $0.35\,\text{ms}$ smaller than that when they stand still. This small difference may be caused by several affecting factors discussed earlier, i.e., human body size and etc. The human body movements increase the variance of $|\epsilon|$, since they create more signal dynamics as seen in Fig.~\ref{fig:raw-samebody-shaking}.

\subsubsection{Independent grounds}

Then, we remove the connection between the two Floras' ground pins, such that they have independent grounds. This setting is consistent with real scenarios, where the wearables are generally not wired. We conduct three tests.

Fig.~\ref{fig:raw-nobody-independent} shows the two Floras' signals when they have no physical contact with any human body. The signals have small oscillations with a frequency of $50\,\text{Hz}$.

\begin{figure}
  \subfigure[Still body]
  {
    \includegraphics{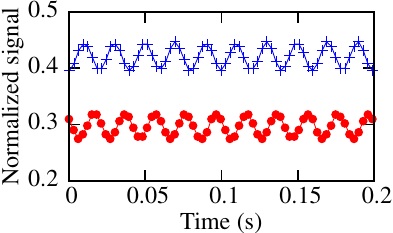}
    \label{fig:raw-samebody-static-independent}
  }
  \subfigure[Moving body]
  {
    \includegraphics{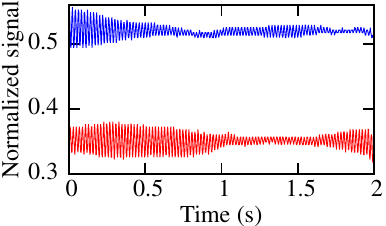}
    \label{fig:raw-samebody-shaking-independent}
  }
  \caption{SEPs on a same wearer (independent grounds).}
  \label{fig:raw-samebody-independent}
\end{figure}

Figs.~\ref{fig:raw-samebody-static-independent} and \ref{fig:raw-samebody-shaking-independent} show the signals of the two Floras worn on two wrists of a researcher when he stands still and walks, respectively.
Compared with the results in Fig.~\ref{fig:raw-samebody-static} based on a shared ground, the two signals in Fig.~\ref{fig:raw-samebody-static-independent} have an offset in their values. This offset is the difference between the electric potentials at the two Floras' grounds. Fig.~\ref{fig:raw-samebody-shaking-independent} shows the signals over two seconds that contain about 100 SEP cycles to better illustrate the changing signal envelopes over time due to the human body movements. Compared with Fig.~\ref{fig:raw-samebody-shaking}, the two signals in Fig.~\ref{fig:raw-samebody-shaking-independent} have different signal envelopes. This is because the electric potentials at the two Floras' grounds, which are also induced by the powerline radiation, are not fully correlated in the presence of human body movements.

\begin{figure}
  \subfigure[Still bodies]
  {
    \includegraphics{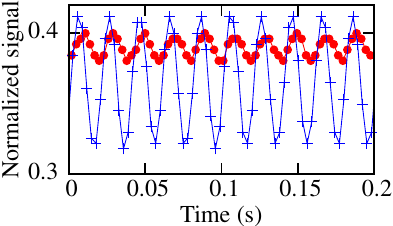}
    \label{fig:raw-diffbody-static-independent}
  }
  \subfigure[Moving bodies]
  {
    \includegraphics{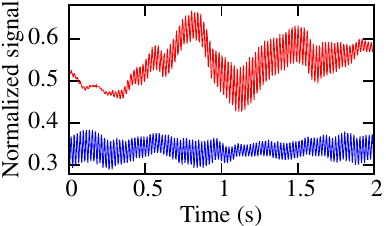}
    \label{fig:raw-diffbody-shaking-independent}
  }
  \caption{SEPs on different wearers (independent grounds).}
  \label{fig:raw-diffbody-independent}
\end{figure}

Figs.~\ref{fig:raw-diffbody-static-independent} and \ref{fig:raw-diffbody-shaking-independent} show the signals of the two Floras worn by two researchers each when they stand still and walk, respectively.
Salient EMR signals can be observed. Moreover, the human movements cause significant fluctuations of DC lines and the signal amplitudes, as seen in Fig.~\ref{fig:raw-diffbody-shaking-independent}.

We also evaluate the synchronism between the two Floras' signals. Fig.~\ref{fig:delay-independent-ground} shows the time displacement's error bars that correspond to the scenarios in Figs.~\ref{fig:raw-samebody-static-independent}, \ref{fig:raw-samebody-shaking-independent}, \ref{fig:raw-diffbody-static-independent}, and \ref{fig:raw-diffbody-shaking-independent}. The average $|\epsilon|$ is below $3\,\text{ms}$. Compared with the results in Fig.~\ref{fig:delay-shared-ground} that are based on a shared ground, the time displacements increase. This is because of the additional uncertainty introduced by the independent floating grounds of the two Floras. Nevertheless, on the same wearer, the average $|\epsilon|$ is about $1\,\text{ms}$ only. The body movements increase the 95\%-percentile of $|\epsilon|$ to $6\,\text{ms}$. In \sect\ref{subsec:conditioning}, we will use a phase-locked loop to reduce the variations of $\epsilon$.

\subsubsection{Summary}
\label{subsubsec:measurement-summary}

From the above measurements, we obtain the following three key observations. First, the human body can act as an antenna that effectively improves the powerline radiation reception. Second, during the human body movements, the SEP amplitude changes. However, the synchronism between the two SEP signals captured by the two nodes on the same wearer or different wearers is still acceptably preserved. In \sect\ref{subsec:conditioning}, we will condition the SEP signals to improve the synchronism. Third, the floating ground of a node introduces additional uncertainty, because the powerline radiation can also generate a varying electric potential at the ground pin. However, the floating ground does not substantially degrade the synchronism between the two nodes' SEP signals. All experiments in the rest of this paper are conducted under the floating ground setting. The above three observations suggest that the SEP induced by powerline radiation is a good periodic signal that can be exploited for synchronizing wearables.

We note that, all the above measurements are conducted in a computer science laboratory that draws electricity from a single power grid phase. Thus, the SEPs received by the Floras in the laboratory have the same phase. Typically, a small area (e.g., a room and an office floor) is supplied by the same power grid phase. Thus, the wearables in the area will sense synchronized SEPs. Two remote wearables may sense different power grid phases. The voltage phase difference will become part of the synchronization error. It is $6.7\,\text{ms}$ and $5.6\,\text{ms}$ in $50\,\text{Hz}$ and $60\,\text{Hz}$ grids, respectively. We will observe this in our experiments presented in \sect\ref{subsec:signal-availability}.

The powerline radiation and SEP are generally unavailable outdoors. Our extensive evaluation in \sect\ref{sec:evaluation} will show the pervasive availability of SEPs in indoor environments. As most of our lifetime is indoors (e.g., 87\% on average for Americans \cite{klepeis2001national}), the SEP will be available for synchronizing wearables.

\section{Design of TouchSync}
\label{sec:approach}

In this section, we present the design of TouchSync. \sect\ref{subsec:overview} overviews the workflow of TouchSync. \sect\ref{subsec:conditioning} presents the signal processing algorithms to generate stable, periodic , and synchronous impulses trains (i.e., Dirac combs) from the SEP signals. \sect\ref{subsec:dc-ntp} presents a synchronization protocol assisted with the Dirac combs. \sect\ref{subsubsec:ambiguity} solves the integer ambiguity problem to complete synchronization.

\subsection{TouchSync Workflow}
\label{subsec:overview}

\begin{figure}
  \includegraphics[width=0.42\textwidth]{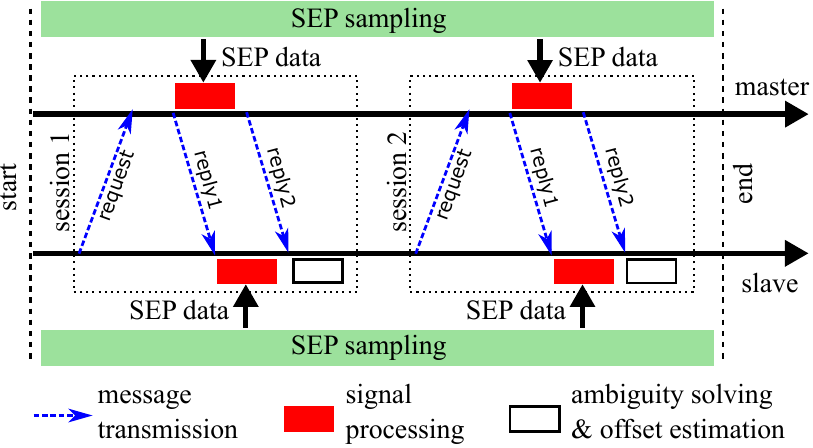}
  \caption{A synchronization process of TouchSync.}
  \label{fig:overview}
\end{figure}

TouchSync synchronizes the clock of a slave to that of a master. This paper focuses on the synchronization between a slave-master pair, which is the basis for synchronizing a network of nodes. A {\em synchronization process}, as illustrated in Fig.~\ref{fig:overview}, is performed periodically or in an on-demand fashion. For instance, the wearer(s) may push some buttons on two wearables to start a synchronization process. The period of the synchronization can be determined by the needed clock accuracy and the clock drift rate. During the synchronization process, both the slave and the master continuously sample the SEP signals and store the timestamped samples into their local buffers.
At the beginning of the synchronization process, the slave node sends a message to the master to signal the start of the sensor sampling.
A synchronization process has multiple {\em synchronization sessions}. Fig.~\ref{fig:overview} shows two sessions. In each session, the slave and the master exchange three messages: \texttt{request}, \texttt{reply1}, and \texttt{reply2}. The \texttt{request} and \texttt{reply1} are used to measure the communication delays. After transmitting the \texttt{reply1}, the master retrieves a segment of SEP signal from its buffer to process and transmits the processing results using the \texttt{reply2} to the slave. Upon receiving the \texttt{reply1}, the slave retrieves a segment of SEP signal from its buffer to process. Upon receiving the \texttt{reply2}, the slave tries to solve the integer ambiguity problem to estimate the offset between the slave's and the master's clocks. If the ambiguity cannot be solved, another synchronization session is initiated; otherwise, the two nodes stop sampling SEPs and the slave uses the estimated offset to adjust its clock and complete the synchronization process.

\subsection{SEP Signal Processing}
\label{subsec:conditioning}

This section presents TouchSync's signal processing illustrated as the filled blocks in Fig.~\ref{fig:overview}. The objective is to generate a highly stable, periodic, and synchronous Dirac comb from a SEP signal with fluctuating DC component and jitters as shown in Figs.~\ref{fig:raw-diffbody-shaking-independent} and \ref{fig:delay-independent-ground}. The algorithms should be compute- and storage-efficient.

\begin{figure}
  \centering
  \includegraphics[width=0.4\textwidth]{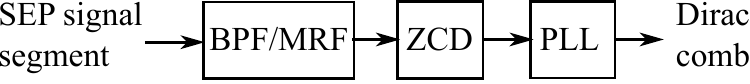}
  \caption{SEP signal processing pipeline.}
  \label{fig:pipeline}
\end{figure}

\begin{figure}
  \begin{minipage}[t]{.235\textwidth}
    \includegraphics{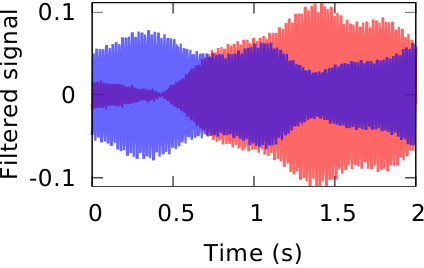}
    \vspace{-2em}
    \caption{BPF output (each color stands for the same node in Fig.~\ref{fig:raw-diffbody-independent}).}
    \label{fig:bpf-result}
  \end{minipage}
  \hspace{0.5em}
  \begin{minipage}[t]{.2\textwidth}
    \centering
    \vspace{-1.1in}
    \includegraphics{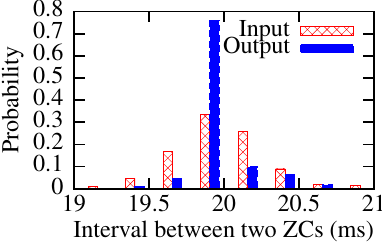}
    \caption{Jitter before and after PLL.}
    \label{fig:jitter}
  \end{minipage}
\end{figure}

We apply a signal processing pipeline illustrated in Fig.~\ref{fig:pipeline}. It has three steps:

\vspace{0.2em}\noindent
{\bf Band-pass filter (BPF) or mean removal filter (MRF):} We apply a 5th-order/6-tap infinite impulse response (IIR) BPF with steep boundaries of a $(45\,\text{Hz}, 55\,\text{Hz})$ passband to remove the fluctuating DC component and high-frequency noises of the SEP signal. The red and blue curves in Fig.~\ref{fig:bpf-result} are the filtering results for the red and blue signals shown in Fig.~\ref{fig:raw-diffbody-shaking-independent}. We can see that the DC components have been removed. For too resource-limited wearables, a MRF that subtracts the running average from the original signal can be used instead of the BPF for much lower compute and storage complexities. Its effect is similar to a low-pass filtering.

\vspace{0.2em}\noindent
{\bf Zero crossing detector (ZCD):} It detects the ZCs, i.e., the time instants when the filtered SEP signal changes from negative to positive. It computes a linear interpolation point between the negative and the consequent positive SEP samples as the ZC to mitigate the impact of low time resolution due to a low SEP sampling rate.

\begin{figure}
  \centering
  \begin{minipage}[t]{.235\textwidth}
    \centering
    \includegraphics{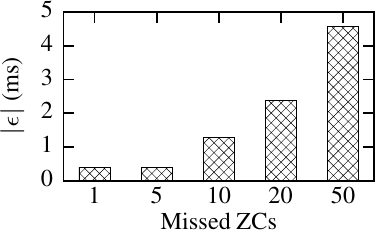}
    \caption{PLL robustness.}
    \label{fig:pll-robustness}
  \end{minipage}
  \begin{minipage}[t]{.235\textwidth}
    \centering
    \includegraphics{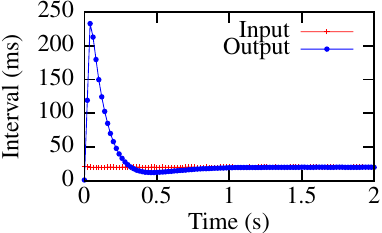}
    \caption{PLL convergence.}
    \label{fig:convergence}
  \end{minipage}
\end{figure}

\vspace{0.2em}\noindent
{\bf Phase-locked loop (PLL):} We apply a software PLL to deal with the ZC jitters and miss detection caused by significant dynamics of the SEP signal. The PLL generates an impulse train using a loop and uses an active proportional integral (PI) controller to tune the interval between two consecutive impulses according to the time differences between the past impulses and the input ZCs. The controller skips the time differences larger than $25\,\text{ms}$ to deal with ZC miss detection. Fig.~\ref{fig:jitter} shows the distributions of the interval between two consecutive ZCs of the PLL's input and output, respectively. The PLL reduces jitters. Before PLL, the minimum and maximum intervals are $13.5\,\text{ms}$ and $30.8\,\text{ms}$, respectively. After PLL, the minimum and maximum intervals are $19.4\,\text{ms}$ and $20.7\,\text{ms}$, respectively. To understand PLL's robustness against ZC miss detection, we artificially drop a number of continuous ZCs to simulate an outage period and evaluate the time displacement $\epsilon$ between the PLL's output and the dropped ZCs. Fig.~\ref{fig:pll-robustness} shows the $|\epsilon|$ versus the number of missed ZCs. When 50 ZCs that last for one second are missed, the caused $|\epsilon|$ is $4.5\,\text{ms}$ only. Moreover, after the outage period, the $|\epsilon|$ restores to sub-ms with five input ZCs only.

Though the above three steps are standard signal processing techniques, they are crucial for TouchSync. Moreover, we tune them to better cater into our needs. In \cite{Rowe2009lowpower}, the PLL is also used to reduce jitters of a clock calibration signal. Different from \cite{Rowe2009lowpower} that continuously runs PLL for continuous clock calibration, TouchSync samples SEP and runs PLL only when clock synchronization is needed. Thus, we configure the PLL to have a short convergence time of about one second, as shown in Fig.~\ref{fig:convergence}. Moreover, owing to a special consideration in the design of TouchSync that will be presented in \sect\ref{subsec:dc-ntp}, the signal processing algorithms operate in an ``offline'' manner, in that they start to work until the whole SEP signal segment to be processed becomes available. This largely simplifies the implementation of these algorithms.

\subsection{NTP Assisted with Dirac Combs}
\label{subsec:dc-ntp}

TouchSync uses the synchronous Dirac combs at the slave and the master to achieve clock synchronization through multiple synchronization sessions. This section presents the protocol and the analysis for a single synchronization session.

\subsubsection{Protocol for a synchronization session}

\begin{figure}
  \centering
  \includegraphics[width=.42\textwidth]{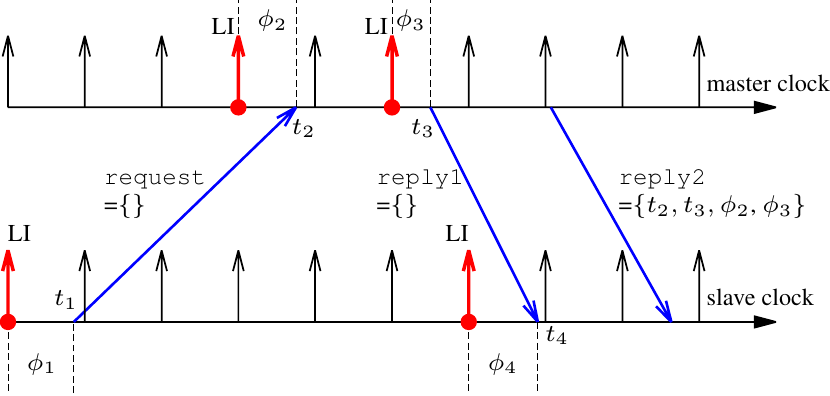}
  \caption{A synchronization session of TouchSync. The vertical arrows represent the impulses of the Dirac combs generated from the SEP signal.}
  \label{fig:touchsync-session}
\end{figure}

A synchronization session of TouchSync is illustrated in Fig.~\ref{fig:touchsync-session}. We explain it from the following two aspects.

\begin{description}

\item[Message exchange and timestamping:]
The session consists of the transmissions of three messages: \texttt{request}, \texttt{reply1}, and \texttt{reply2}. The \texttt{request} and \texttt{reply1} messages are similar to the two UDP packets used by NTP. Their transmission and reception timestamps, i.e., $t_1$, $t_2$, $t_3$, and $t_4$, are obtained upon the corresponding messages are passed/received to/from the OS, as illustrated in Fig.~\ref{fig:timestamping}.

The master will transmit the auxiliary \texttt{reply2} message to convey the results of its signal processing, which is detailed below.

\item[Signal processing and clock offset estimation:]
After the master has transmitted the \texttt{reply1} message, the master (i) retrieves from its signal buffer a SEP signal segment that covers the time period from $t_2$ to $t_3$ with some safeguard ranges before $t_2$ and after $t_3$, (ii) feeds the signal processing pipeline in \sect\ref{subsec:conditioning} with the retrieved SEP signal segment to produce a Dirac comb as illustrated in Fig.~\ref{fig:touchsync-session}, and (iii) identifies the last impulses (LIs) in its Dirac comb that are right before the time instants $t_2$ and $t_3$, respectively. The LIs are illustrated by thick red arrows in Fig.~\ref{fig:touchsync-session}. Then,  the master computes the elapsed times from $t_2$'s LI to $t_2$ and $t_3$'s LI to $t_3$, which are denoted by $\phi_2$ and $\phi_3$, respectively. The $\phi_2$ and $\phi_3$ are the {\em phases} of the $t_2$ and $t_3$ with respect to the Dirac comb. After that, the master transmits the \texttt{reply2} message that contains $t_2$, $t_3$, $\phi_2$, and $\phi_3$ to the slave. After receiving the \texttt{reply1} message, the slave retrieves a SEP signal segment that covers the time period from $t_1$ to $t_4$ with some safeguard ranges, executes the signal processing pipeline, identifies the LIs right before $t_1$ and $t_4$, and computes the phases $\phi_1$ and $\phi_4$, as illustrated in Fig.~\ref{fig:touchsync-session}. After receiving the \texttt{reply2} message, based on $\{t_1, t_2, t_3, t_4\}$ and $\{\phi_1, \phi_2, \phi_3, \phi_4\}$, the slave uses the approach in \sect\ref{subsubsec:offset-analysis} to analyze the offset between the slave's and master's clocks. From the PLL convergence speed shown in Fig.~\ref{fig:convergence}, we set the safeguard to one second.
\end{description}

Now, we discuss several design considerations for the protocol described above. TouchSync uses the \texttt{request} and \texttt{reply1} messages to measure the clock offset, while the \texttt{reply2} is an auxiliary message to convey the timestamps $t_2$, $t_3$ and the measurements $\phi_2$, $\phi_3$. With this auxiliary message, we can decouple the task of timestamping the reception of \texttt{request} and the transmission of \texttt{reply1} from the signal processing task of generating the Dirac comb and computing $\phi_2$ and $\phi_3$. On many platforms (e.g., Android Wear and watchOS), continuously sampled sensor data is passed to the application block by block. With the decoupling, the master can compute $\phi_2$ and $\phi_3$ after the \texttt{reply1} is transmitted and the needed SEP data blocks become available. This is why the signal processing algorithms in \sect\ref{subsec:conditioning} can operate in an ``offline'' manner.

\subsubsection{Clock offset analysis}
\label{subsubsec:offset-analysis}

We now analyze the offset between the slave's and the master's clocks based on $\{t_1$, $t_2$, $t_3$, $t_4\}$ and $\{\phi_1$, $\phi_2$, $\phi_3$, $\phi_4\}$. Denote by $T$ the period of the Dirac comb. In our region served by a $50\,\text{Hz}$ grid, the nominal value for $T$ is $20\,\text{ms}$. To capture the small deviation from the nominal value, $T$ can be also easily computed as the average interval between consecutive impulses of the Dirac comb. We define the {\em rounded phase differences} $\theta_{q}$ and $\theta_{p}$ (which correspond to the \texttt{re\underline{q}uest} and \texttt{re\underline{p}ly1} messages, respectively) as
\begin{equation}
\theta_{q} = \left\{
\begin{array}{ll}
\phi_2 - \phi_1, & \text{if } \phi_2 - \phi_1 \ge 0;\\
\phi_2 - \phi_1 + T, & \text{otherwise}.
\end{array}
\right.
\label{eq:thetaq}
\end{equation}
\begin{equation}
\theta_{p} = \left\{
\begin{array}{ll}
\phi_4 - \phi_3, & \text{if } \phi_4 - \phi_3 \ge 0;\\
\phi_4 - \phi_3 + T, & \text{otherwise}.
\end{array}
\right.
\label{eq:thetap}
\end{equation}
As $\phi_k$ is the elapsed time from $t_k$'s LI to $t_k$, we have $0 \le \phi_k < T$, for $k \in [1, 4]$. From Eqs.~(\ref{eq:thetaq}) and (\ref{eq:thetap}), we can verify that $0 \le \theta_q < T$ and $0 \le \theta_p < T$. From our measurements in \sect\ref{subsec:ntp-perf}, the times for transmitting the \texttt{request} and \texttt{reply1} messages can be longer than $T$. Thus, we use $i$ to denote the non-negative integer number of the Dirac comb's periods elapsed from the time of sending \texttt{request} to the time of receiving it at the master, and $j$ to denote the non-negative integer number of the Dirac comb's periods elapsed from the time of sending \texttt{reply1} to the time of receiving it at the slave.

We denote $\tau_q$ and $\tau_p$ the actual times for transmitting the \texttt{re\underline{q}uest} and the \texttt{re\underline{p}ly} messages, respectively. Thus,
\begin{equation}
\tau_q = \theta_q + i \cdot T - \epsilon, \qquad \tau_p = \theta_p + j \cdot T + \epsilon,
\label{eq:taus}
\end{equation}
where $\epsilon$ is the time displacement between the slave's and master's Dirac combs. Here, we assume a constant $\epsilon$ to simplify the discussion. Therefore, the RTT computed by $\mathrm{RTT} = (t_4 - t_1) - (t_3 - t_2)$ must satisfy
\begin{equation}
\mathrm{RTT} = \tau_q + \tau_p = \theta_q + \theta_p + (i + j) \cdot T.
\label{eq:RTT-theta}
\end{equation}
In Eq.~(\ref{eq:RTT-theta}), $\mathrm{RTT}$, $\theta_q$, and $\theta_p$ are measured in the synchronization session illustrated in Fig.~\ref{fig:touchsync-session}; $i$ and $j$ are unknown non-negative integers. If the $i$ and $j$ can be determined, the estimated offset between the slave's and the master's clocks, denoted by $\delta$, can be computed by either one of the following formulas:
\begin{equation}
\delta = t_1 - (t_2 - \tau_q) = t_1 - t_2 + \theta_q + i \cdot T - \epsilon,
\end{equation}
\begin{equation}
\delta = t_4 - (t_3 + \tau_p) = t_4 - t_3 - \theta_p - j \cdot T - \epsilon.
\label{eq:touchsync-offset}
\end{equation}
It can be easily verified that the above two formulas give the same result. The analysis in the rest of this paper chooses to use Eq.~(\ref{eq:touchsync-offset}).
The $\epsilon$ is generally unknown. If we ignore it in Eq.~(\ref{eq:touchsync-offset}) to compute $\delta$, it becomes part of the clock offset estimation error.

Eq.~(\ref{eq:RTT-theta}) is an {\em integer-domain} underdetermined problem. Clearly, from Eq.~(\ref{eq:RTT-theta}), both $i$ and $j$ belong to the range $\left[ 0, \frac{\mathrm{RTT} - \theta_q - \theta_p}{T} \right]$. Thus, Eq.~(\ref{eq:RTT-theta}) has a finite number of solutions for $i$ and $j$. Note that, under the original NTP principle, we have a {\em real-domain} underdetermined problem of $\mathrm{RTT} = \tau_q + \tau_p$ that has infinitely many solutions. NTP chooses a solution by assuming $\tau_q=\tau_p$, which does not hold in general. Thus, by introducing the Dirac combs, the {\em ambiguity} in determining $\tau_q$, $\tau_p$, and $\delta$ is substantially reduced from infinitely many possibilities to finite possibilities. Though we still have ambiguity in the integer domain, our analysis and extensive numeric results in \sect\ref{subsubsec:ambiguity} show that the ambiguity can be solved.

Note that, in \cite{dima17}, the periodic and synchronous power grid voltage signals collected directly from power outlets are used to synchronize two nodes that have high-speed wired network connections. The approach in \cite{dima17} also uses the elapsed times from LIs (i.e., $\phi_1$, $\phi_2$, $\phi_3$, $\phi_4$) to deal with asymmetric communication delays and improve synchronization accuracy. However, due to the high-speed connectivity, it only considers the case where both $i$ and $j$ are zero. In contrast, with wireless connectivity, $i$ and $j$ are random and often non-zero due to the access time (cf.~\sect\ref{subsec:ntp}). Estimating $i$ and $j$ is challenging and it is the subject of \sect\ref{subsubsec:ambiguity}.

\subsection{Integer Ambiguity Solver (IAS)}
\label{subsubsec:ambiguity}

Before we present the approach to solving integer ambiguity, we make the following two assumptions for simplicity of exposition. First, we assume that the ground-truth clock offset $\delta_{GT}$ is a constant during a synchronization process. From our performance evaluation in \sect\ref{sec:evaluation}, TouchSync generally takes less than one second to achieve synchronization. Typical crystal oscillators found in MCUs and personal computers have drift rates of 30 to 50 ppm \cite{hao2011wizsync}. Thus, the maximum drift of the offset between two clocks during one second is $50\,\text{ppm} \times 1\,\text{s} \times 2 = 0.1\,\text{ms}$. This drift is smaller than the ms-level time displacement $\epsilon$ between two SEP signals, which dominates the synchronization error of TouchSync. Second, we assume $\epsilon=0$. In \sect\ref{subsubsec:ambiguity-discussion}, we will discuss how to deal with non-zero and time-varying $\epsilon$.

We let $i_{\min}$ and $i_{\max}$ denote the minimum and maximum possible values for $i$; $j_{\min}$ and $j_{\max}$ denote the minimum and maximum possible values for $j$. For instance, from our one-way message transmission time measurements summarized in Fig.~\ref{fig:one-way-delays}, the BLE's slave-to-master transmission times are always greater than $30\,\text{ms}$. Thus, we may set $i_{\min} = 1$, since in our region $T$ is $20\,\text{ms}$. When we have no prior knowledge about the ranges for $i$ and $j$, we may simply set $i_{\min} = j_{\min} = 0$ and $i_{\max} = j_{\max} = \frac{\mathrm{RTT} - \theta_q - \theta_p}{T}$.

\sect\ref{subsubsec:ambiguity-discussion} will discuss how the use of the prior knowledge impacts on the integer ambiguity solving.

TouchSync performs multiple synchronization sessions to solve the integer ambiguity problem. In this section, we use $x[k]$ to denote a quantity $x$ in the $k^\text{th}$ synchronization session. For instance, $\mathrm{RTT}[k]$ denotes the measured RTT in the $k^\text{th}$ session. From Eqs.~(\ref{eq:RTT-theta}) and (\ref{eq:touchsync-offset}), for the $k^\text{th}$ synchronization session, we have
\begin{equation}
\left\{
\begin{array}{l}
\mathrm{RTT}[k] = \theta_{q}[k] + \theta_{p}[k] + (i[k] + j[k]) \cdot T; \\
\delta = t_4[k] - t_3[k] - \theta_{p}[k] - j[k] \cdot T; \\
i_{\min} \le i[k] \le i_{\max}, \quad j_{\min} \le j[k] \le j_{\max}.
\end{array}
\right.
\label{eq:ambiguity-equation}
\end{equation}
If TouchSync performs $K$ synchronization sessions, we have an underdetermined system of $2K$ equations with $(2K+1)$ unknown variables (i.e., $\delta$ and $\{i[k], j[k] | k \in [1, K]\}$). In the integer domain, such an underdetermined system can have a unique solution.

\subsubsection{An example of unique solution}

\begin{figure}
  \includegraphics{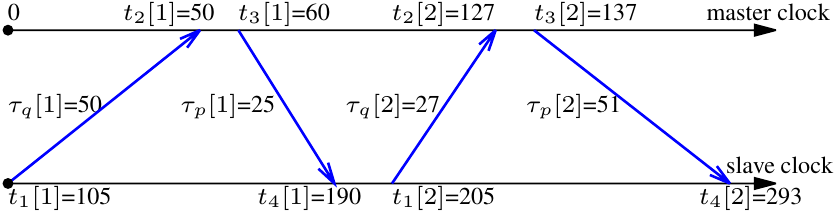}
  \caption{An example of solving the integer ambiguity. The transmissions of the auxiliary \texttt{reply2} messages are omitted in the illustration.}
  \label{fig:integer-ambiguity}
\end{figure}

We use an example in Fig.~\ref{fig:integer-ambiguity} to illustrate. The unit for time is ms, which is omitted in the following discussion for conciseness. In this example, $T=20$, $i_{\min}= j_{\min} = 1$, $i_{\max} = j_{\max} =4$, and the ground-truth clock offset $\delta_{GT} = 105$. Two synchronization sessions are performed. The timestamps and the actual message transmission delays are shown in Fig.~\ref{fig:integer-ambiguity}. The ground-truth values for $i$ and $j$ in the two synchronization sessions are: $i[1]=2$, $j[1]=1$, $i[2]=1$, and $j[2]=2$. The RTTs can be computed as $\mathrm{RTT}[1] = 75$ and $\mathrm{RTT}[2] = 78$. With any synchronous Dirac combs, from Eqs.~(\ref{eq:thetaq}) and (\ref{eq:thetap}), the rounded phase differences computed by the two nodes must be $\theta_q[1] = 10$, $\theta_p[1] = 5$, $\theta_q[2]=7$, and $\theta_p[2] = 11$. For the first synchronization session, Eq.~(\ref{eq:ambiguity-equation}) has two possible solutions only:
\begin{equation}
\{i[1] \!=\! 1, j[1] \!=\! 2, \delta \!=\! 85\}, \quad \{i[1] \!=\! 2, j[1] \!=\! 1, \delta \!=\! 105\}.
\label{eq:solution1}
\end{equation}
For the second synchronization session, Eq.~(\ref{eq:ambiguity-equation}) has two possible solutions only as well:
\begin{equation}
\{i[2] \!=\! 1, j[2] \!=\! 2, \delta \!=\! 105\}, \quad \{i[2] \!=\! 2, j[2] \!=\! 1, \delta \!=\! 125 \}.
\label{eq:solution2}
\end{equation}
From Eqs.~(\ref{eq:solution1}) and (\ref{eq:solution2}), $\delta=105$ is the only common solution. Thus, we conclude that $\delta$ must be 105.

\subsubsection{Program for solving integer ambiguity}
\label{subsubsec:program}

From the above example, due to the diversity of the ground-truth values of $i$ and $j$ in multiple synchronization sessions, the intersection of the $\delta$ solution spaces of these synchronization sessions can be a single value. Thus, the integer ambiguity problem is solved. On the contrary, if the ground-truth $i$ and $j$ do not change over multiple synchronization sessions, the ambiguity remains. As explained in \sect\ref{subsec:ntp}, with application-layer timestamping, the message transmission times are highly dynamic due to the uncertain OS overhead and MAC. Such uncertainties and dynamics, which are undesirable in the original theme of NTP, interestingly, become desirable for solving the integer ambiguity in TouchSync.

From the above key observation, TouchSync performs the synchronization session illustrated in Fig.~\ref{fig:touchsync-session} {\em repeatedly} until the intersection among the $\delta$ solution spaces of all the synchronization sessions converges to a single value. Algorithms~\ref{alg:slave} and \ref{alg:master} provide the pseudocode for TouchSync's slave and master programs.

\renewcommand{\algorithmicrequire}{\textbf{Given:}}
\renewcommand{\algorithmiccomment}[1]{// #1}

\begin{algorithm}
  \caption{Slave's pseudocode for a synchronization process}
\label{alg:slave}
\renewcommand{\algorithmicwhile}{\textbf{event}}
\renewcommand{\algorithmicfor}{\textbf{command}}
\begin{algorithmic}[1]
\small

\STATE Global variables: $t_1$, $t_4$, $\delta$'s solution space $\Delta = \varnothing$, session index $k = 0$

\STATE

\FOR{start\_sync\_session()}
\STATE $k = k + 1$
\STATE \texttt{$t_1$} = \texttt{read\_system\_clock()}
\STATE send message \texttt{request} = \{ \} to master
\ENDFOR

\STATE

\WHILE{\texttt{reply1} received from master}
\STATE \texttt{$t_4$} = \texttt{read\_system\_clock()}
\STATE wait until SEP data covering $t_1$ and $t_4$ are available
\label{on-receive-reply1-start}
\STATE run the SEP signal processing pipeline in \sect\ref{subsec:conditioning}
\STATE compute $\phi_1$ and $\phi_4$
\label{on-receive-reply1-end}
\ENDWHILE

\STATE

\WHILE{\texttt{reply2} received from master}
\STATE compute $\theta_q$ and $\theta_p$ using Eqs.~(\ref{eq:thetaq}) and (\ref{eq:thetap}), respectively
\label{on-receive-reply2-start}
\STATE $\mathrm{RTT} = (t_4 - t_1) - (\text{reply2}.t_3 - \text{reply2}.t_2)$
\STATE solve Eq.~(\ref{eq:ambiguity-equation}), $\Delta'$ denotes the set of all possible solutions for $\delta$
\STATE {\bf if} $k==1$: $\Delta = \Delta'$; {\bf else}: $\Delta = \Delta \cap \Delta'$
\label{line:intersection}
\STATE {\bf if} $\Delta$ has only one element $\delta$: use $\delta$ to adjust clock;
\STATE {\bf else}: start\_sync\_session() \COMMENT{start a new synchronization session}
\ENDWHILE

\end{algorithmic}
\end{algorithm}

\begin{algorithm}
\caption{Master's pseudocode for a synchronization process}
\label{alg:master}
\renewcommand{\algorithmicwhile}{\textbf{event}}
\renewcommand{\algorithmicfor}{\textbf{command}}
\begin{algorithmic}[1]
\small

\WHILE{\texttt{request} received from slave}
\STATE \texttt{$t_2$} = \texttt{read\_system\_clock()}
\STATE ... \COMMENT{execute other compute tasks}
\STATE \texttt{$t_3$} = \texttt{read\_system\_clock()}
\STATE send message \texttt{reply1} = \{ \} to slave
\STATE wait until SEP data covering $t_2$ and $t_3$ are available
\label{send-reply1-done-start}
\STATE run the SEP signal processing pipeline in \sect\ref{subsec:conditioning}
\STATE compute $\phi_2$ and $\phi_3$
\label{send-reply1-done-end}
\STATE send message \texttt{reply2} = \{$t_2$, $t_3$, $\phi_2$, $\phi_3$\} to slave
\ENDWHILE
\end{algorithmic}
\end{algorithm}

\subsubsection{Convergence speed}
\label{subsubsec:convergence-speed}

We run a set of numeric experiments to understand the convergence speed of the IAS. We use the number of synchronization sessions until convergence to characterize the convergence speed, which is denoted by $K$ in the rest of this paper. We fix $i_{\min}$ and $j_{\min}$ to be zero. For a certain setting $\langle i_{\max}, j_{\max} \rangle$, we conduct 100,000 synchronization processes to assess the distribution of $K$. For each synchronization session of a synchronization process, we randomly and uniformly generate the ground-truth $i$ and $j$, as well as $\theta_q$ and $\theta_p$ within their respective ranges, i.e., $i \in [0, i_{\max}]$, $j \in [0, j_{\max}]$, and $\theta_q, \theta_p \in [0, T)$. Then, we simulate the integer ambiguity solving program presented in \sect\ref{subsubsec:program} to measure the $K$ for each synchronization process. 
In practice, the $i$, $j$, $\theta_q$ and $\theta_p$ may not follow the uniform distributions. But the numeric results here help us understand the convergence speed. In Section~\ref{subsec:signal-availability}, we will evaluate the convergence speed in real-world settings.

\begin{figure}
  \subfigure[Average $K$ vs. $i_{\max}$ and $j_{\max}$.]
  {
    \includegraphics{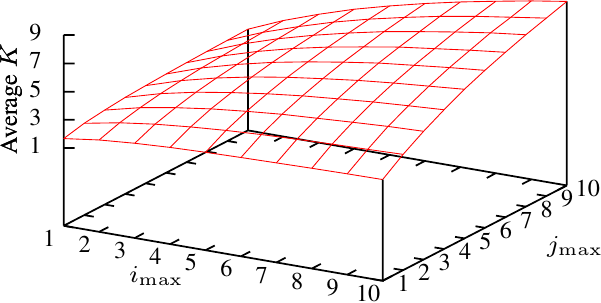}
    \label{fig:convergence1}
  }

  \subfigure[Box plot for $K$ (box represents the 1st and 3rd quartiles; whiskers represent minimum and maximum values.)]
  {
    \includegraphics{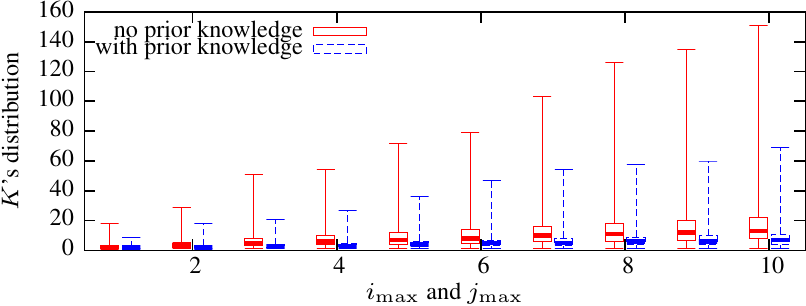}
    \label{fig:convergence2}
  }
  \caption{Convergence speed of IAS.}
\end{figure}

In Fig.~\ref{fig:convergence1}, each grid point is the average of all $K$ values in the 100,000 synchronization processes under a certain $\langle i_{\max}, j_{\max} \rangle$ setting. Fig.~\ref{fig:convergence2} shows the box plot for $K$ under each setting where $i_{\max}=j_{\max}$. We note that all simulated synchronization processes converge. From the two figures, even if $i_{\max} = j_{\max} = 10$ (which means that the one-way communication delays are up to $200\,\text{ms}$ for $T=20\,\text{ms}$), the average $K$ is nine only. Although the $K$'s distribution has a long tail as shown in Fig.~\ref{fig:convergence2}, 75\% of the $K$ values are below 11. This result is consistent with our real experiment results in Table~\ref{tab:indoor-environments} of \sect\ref{subsec:signal-availability}, where most $K$ values are two only and the largest $K$ is 12.

\subsubsection{Discussions}
\label{subsubsec:ambiguity-discussion}

First, we discuss how to address non-zero and time-varying $\epsilon$. From the analysis in Eq.~(\ref{eq:ambiguity-equation}) that is based on $\epsilon=0$, the difference between two $\delta$ solutions is multiple of $T$. This can also be seen from Eqs.~(\ref{eq:solution1}) and (\ref{eq:solution2}). In practice, $\epsilon$ can be non-zero and time-varying. It will be a major part of the $\delta$ estimation error. From Fig.~\ref{fig:delay-independent-ground}, the $|\epsilon|$ is at most $6\,\text{ms}$. Thus, the resulted variation to the $\delta$ solutions will be less than a half and one third of $T$, in the regions served by $60\,\text{Hz}$ and $50\,\text{Hz}$ power grids, respectively. Therefore, we can still correctly identify the correspondence among the $\delta$ elements in the set intersection operation in Line~\ref{line:intersection} of the slave's program in Algorithm~\ref{alg:slave}. Specifically, if two $\delta$ elements have a difference smaller than $T/2$, they should be considered the same element in the set intersection operation; otherwise, they are different elements. 
For this correspondence identification to be correct, the $\epsilon$ needs to be smaller than $T/2$.
After convergence, the final $\delta$ can be computed as the average of the $\delta$ elements that are considered the same. We have incorporated this in our implementation of TouchSync.

Second, we discuss how the use of the prior knowledge (i.e., $i_{\min}$, $i_{\max}$, $j_{\min}$, and $j_{\max}$) impacts on the integer ambiguity solving. With the prior knowledge, we may shrink the search range for $i$ and $j$ to speed up the convergence of the IAS. 
The prior knowledge can be based on the statistical information obtained in offline experiments.
For instance, a group of the box plots are the results for the IAS with the prior knowledge of $i_{\max}$ and $j_{\max}$. The IAS can search the $i$ and $j$ within the ranges of $[0, i_{\max}]$ and $[0, j_{\max}]$, respectively. The other group of results are for the IAS without the prior knowledge. Thus, the IAS has to search within the range of $\left[ 0, \frac{\mathrm{RTT} - \theta_q - \theta_p}{T} \right]$ for both $i$ and $j$. We can see that, if no prior knowledge is used, the $K$ increases. But the IAS still always converges.
Once the IAS converges, the synchronization error of TouchSync mainly depends on the time displacement $\epsilon$.

\section{Implementation of TouchSync}
\label{sec:implementation}

\subsection{\texttt{touchsync.h} Header and Implementations}

Designed as an application-layer clock synchronization approach, TouchSync can be implemented as an app or part of an app, purely based on the standard wearable OS calls to sample the SEP signal, exchange network messages, and timestamp them in the application layer. To simplify the adoption of TouchSync by application developers, we have implemented TouchSync's platform-independent tasks (i.e., SEP signal processing and IAS) in ANSI C and provide them in a \texttt{touchsync.h} header file \cite{touchsync-h}. As most embedded and IoT platforms are C compatible, our C implementation is applicable to a wide range of wearables. Wrappers for other programming languages can also be implemented. The header file defines a circular buffer to store the SEP signal. It provides four functions to be used by the application developer:\vspace{.2em}\\
$\bullet$ \texttt{buf\_add()} pushes a new SEP sample to the circular buffer;\\
$\bullet$ \texttt{on\_receive\_reply1()} implements Line~\ref{on-receive-reply1-start}-\ref{on-receive-reply1-end} of Algorithm~\ref{alg:slave};\\
$\bullet$ \texttt{on\_receive\_reply2()} implements Line~\ref{on-receive-reply2-start}-\ref{line:intersection} of Algorithm~\ref{alg:slave};\\
$\bullet$ \texttt{send\_reply1\_done()} implements Line~\ref{send-reply1-done-start}-\ref{send-reply1-done-end} of Algorithm~\ref{alg:master}.

Other tasks of TouchSync, i.e., sensor sampling, synchronization message exchange and timestamping, are platform dependent. We leave them for the application developer to implement. As these tasks are basics for embedded programming, by using the four functions provided by \texttt{touchsync.h}, application developers without much knowledge in signal processing can readily implement TouchSync on different platforms. Our own Arduino and TinyOS programs that implement TouchSync's work flow have about 50 and 150 lines of code only, respectively.

\subsection{Benchmarking}

To understand the overhead of TouchSync, we deploy our TinyOS and Arduino implementations to Zolertia's Z1 motes \cite{z1} and Floras, respectively. The Z1 mote is equipped with an MSP430 MCU ($1\,\text{MHz}$, $8\,\text{KB}$ RAM) and a CC2420 802.15.4 radio. Both implementations sample SEP at $333\,\text{Hz}$. On Z1, we configure the length of the circular buffer defined in \texttt{touchsync.h} to be 512. Thus, this circular buffer can store 1.5 seconds of SEP data. This is sufficient, because the time periods $[t_2, t_3]$ and $[t_1, t_4]$ that should be covered by the SEP signal segments to be retrieved from the circular buffer and processed by the master and slave are generally a few ms and below $100\,\text{ms}$, respectively. On Flora, we configure the circular buffer length to be 400 and redefine its data type such that TouchSync can fit into Flora's limited RAM space of $2.5\,\text{KB}$. Table~\ref{tab:overhead} tabulates the memory usage of TouchSync and the computation time of different processing tasks. On Z1, a total of $421\,\text{ms}$ processing time is needed for a synchronization session. The BPF uses a major portion of the processing time. Flora cannot adopt BPF because of RAM shortage. It uses MRF instead, which consumes much less RAM and processing time.

Z1 and Flora are two representative resource-constrained platforms. The successful implementations of TouchSync on them suggest that TouchSync can also be readily implemented on other more resourceful platforms.

\begin{table}
  \caption{Storage and compute overhead of TouchSync.}
  \label{tab:overhead}
  \begin{tabular}{c|cc|cccc}
    \hline
    \multirow{2}{*}{\bf Platform} & \multicolumn{2}{c|}{{\bf Memory use} (KB)} & \multicolumn{4}{c}{{\bf Processing time} (ms)} \\
    \cline{2-7}
    & ROM & RAM & BPF/MRF & ZCD & PLL & IAS \\
    \hline
    Z1 & 10 & 5 & 364 & 9 & 48 & 1 \\
    Flora & 17 & 1.9* & 1.3 & 3 & 15 & 0.8 \\
    \hline
  \end{tabular}
  \begin{minipage}[t]{0.5\textwidth}
    \small
  * Estimated based on buffer lengths.
  \end{minipage}
\end{table}
\section{Performance Evaluation}
\label{sec:evaluation}

We conduct extensive experiments to evaluate the performance of TouchSync in various real environments. Each experiment uses two Flora nodes, which act as the TouchSync slave and master, respectively. As Flora does not support BLE master mode, the two Floras cannot communicate directly. Thus, we use a RPi that operates as a BLE master to relay the data packets between the two Floras. This setting is consistent with most body-area networks with a smartphone as the hub. If the hub can also sample powerline radiation or SEP\footnote{Our preliminary experiments show that a smartphone can capture the powerline radiation by sampling its built-in microphone and then applying a BPF on the collected audio data. Thus, it is possible to implement TouchSync on smartphones.
}, each wearable can also synchronize with the hub directly using TouchSync.
We use the approach discussed in \sect\ref{subsec:hardware} to obtain the ground truth clock of each Flora.
The details and the results of our experiments are presented below.

\subsection{Signal Strength and Wearing Position}

\begin{figure}
  \begin{minipage}[t]{.23\textwidth}
    \includegraphics{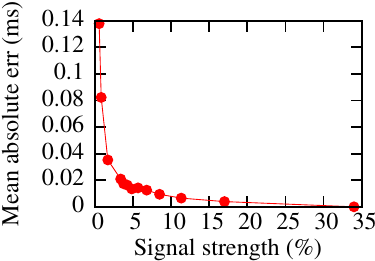}
    \caption{Impact of signal strength.}
    \label{fig:signal-strength}
  \end{minipage}
  \hspace{0.3em}
  \begin{minipage}[t]{.23\textwidth}
    \includegraphics{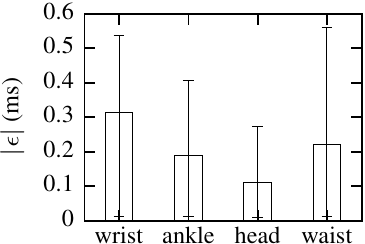}
    \caption{Impact of wearing position.}
    \label{fig:body-position}
  \end{minipage}
\end{figure}

As the intensity of powerline radiation attenuates with distance, SEPs will have varying signal strength. Thus, we evaluate the impact of the SEP signal strength on the performance of the signal processing pipeline in \sect\ref{subsec:overview}. We measure the signal strength as follows. For a full-scale sinusoid signal with a peak-to-peak amplitude of one (normalized using ADC's reference voltage), its standard deviation is $0.5/\sqrt{2}=0.354$. The signal strength of a normalized sinusoid with a standard deviation of $\sigma$ is defined as $\sigma / 0.354$. Thus, a 100\% signal strength suggests a full-scale signal for the ADC. For this experiment, we use a Flora to record a SEP signal. The strength of this signal is 34\%. We feed the signal processing pipeline with this signal to generate a series of {\em baseline} ZCs. Then, we scale down the amplitude of this signal, re-quantize it, and process it using the pipeline to generate another series of ZCs. We use the mean absolute error (MAE) of these ZCs with respect to the baseline ZCs as the error metric. Fig.~\ref{fig:signal-strength} shows the MAE versus the strength of the scaled down signal. When we scale down the signal by 60 times, yielding a signal strength of 0.6\%, the ZCs' MAE is $0.14\,\text{ms}$ only. This suggests that TouchSync can still detect the ZCs accurately even when the SEP signal is rather weak. TouchSync can inform the wearer if the signal is too weak to detect ZCs reliably.

We evaluate the impact of the wearing position on the synchronism of SEP signals. A researcher wears a Flora on his left wrist. Then, he conducts four tests by fixing the second Flora to his right wrist, right ankle, forehead, and waist, respectively. 
Each test lasts for two minutes.
Fig.~\ref{fig:body-position} shows the error bars (5\%-95\% confidence interval) for the absolute time displacement $|\epsilon|$ between the two Floras in these four tests. The average $|\epsilon|$ values in the four tests are close. This suggests that the wearing positions have little impact on the synchronism of SEP signals and the synchronization accuracy of TouchSync.

\subsection{Evaluation in Various Environments}
\label{subsec:signal-availability}

We evaluate the SEP signal strength and the accuracy of TouchSync in various indoor environments.

\begin{figure}
  \centering
  \includegraphics[width=.4\textwidth]{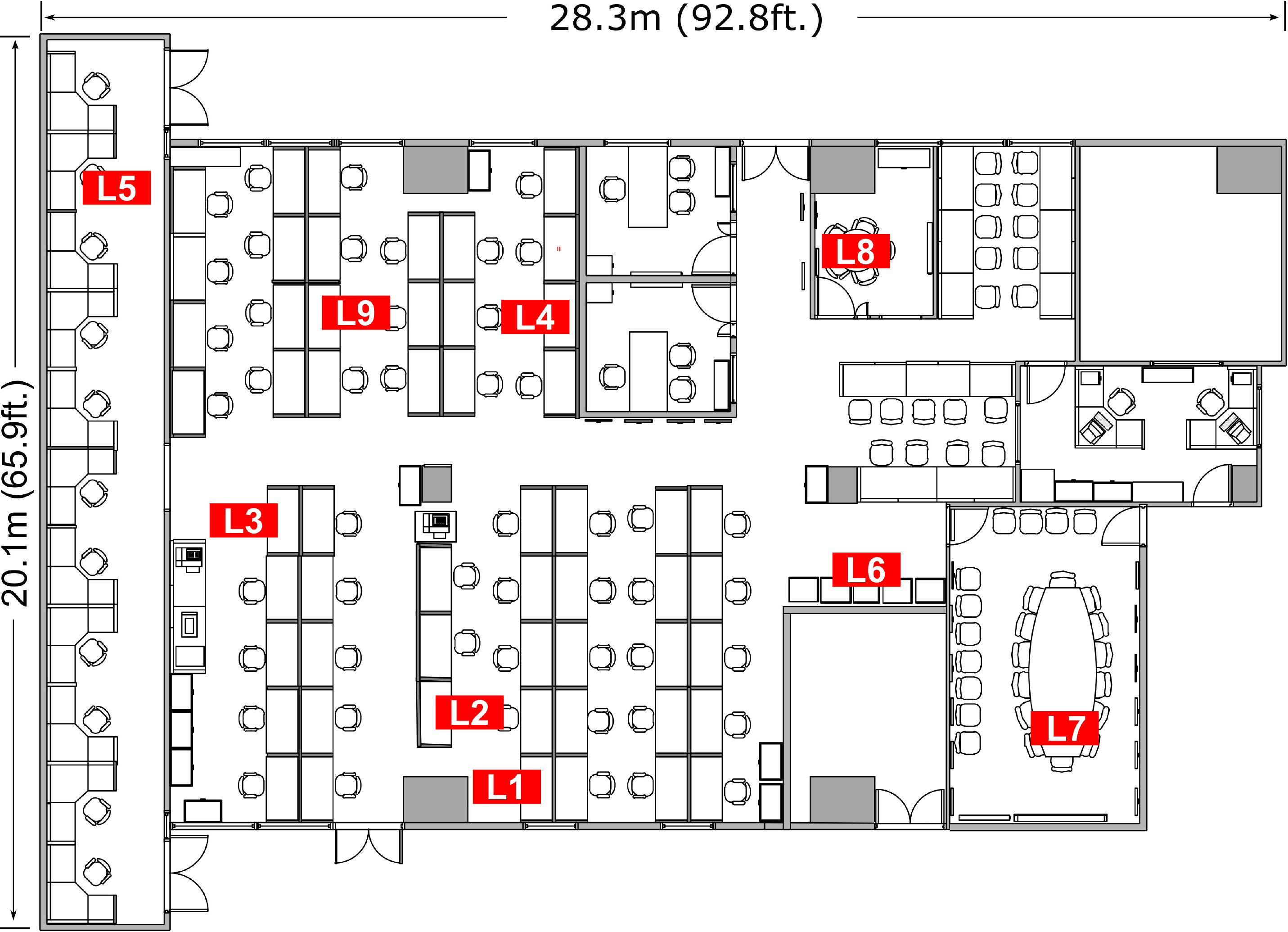}
  \caption{Laboratory floor plan with test points marked.}
  \label{fig:cncl}
\end{figure}

\begin{table}
	\caption{Signal strength and TouchSync accuracy.}
        \label{tab:indoor-environments}
	\begin{tabular}{c|c|ccc|ccc}
          \hline
		& {\bf Test} & \multicolumn{3}{c|}{\bf Without skin contact} & \multicolumn{3}{c}{\bf With skin contact} \\
		\cline{3-8}
		& {\bf point} & Signal & $K$ & error & Signal & $K$ & error \\
		& & strength & & (ms) & strength & & (ms) \\
		\hline
\multirow{9}{*}{\rotatebox[origin=c]{90}{\bf Laboratory}} & L1	&	2.6\%	&	3	&	-0.2	&	84.7\%	&	2	&	-0.7	\\ 
& L2	&	3.2\%	&	2	&	 0.3	&	31.5\%	&	3	&	-0.7	\\ 
& L3	&	2.3\%	&	2	&	-2.5	&	26.1\%	&	2	&	 0.5	\\ 
& L4	&	4.0\%	&	1	&	-0.6	&	33.7\%	&	2	&	0.0	\\ 
& L5	&	0.8\%	&	15	&	 1.1	&	3.3\%	&	2	&	-0.2	\\ 
& L6	&	5.7\%	&	10	&	-0.4	&	39.5\%	&	10	&	-0.0	\\ 
& L7	&	2.3\%	&	n.a.	&	n.a.	&	3.0\%	&	2	&	-0.9	\\ 
& L8	&	4.6\%	&	2	&	-1.5	&	8.3\%	&	2	&	 0.6	\\ 
& L9	&	2.6\%	&	1	&	-1.2	&	67.4\%	&	2	&	-0.9	\\ 
\hline
\multirow{9}{*}{\rotatebox[origin=c]{90}{\bf Home}} & H1	&	4.2\%	&	2	&	-1.1	&	8.9\%	&	2	&	-0.8	\\ 
& H2	&	3.4\%	&	1	&	-0.9	&	14.5\%	&	2	&	-1.0	\\ 
& H3	&	4.6\%	&	1	&	-1.3	&	44.9\%	&	2	&	 0.2	\\ 
& H4	&	7.8\%	&	n.a.	&	n.a.	&	39.2\%	&	2	&	 0.3	\\ 
& H5	&	3.8\%	&	1	&	-1.6	&	3.9\%	&	1	&	-2.8	\\ 
& H6	&	3.9\%	&	4	&	-4.4	&	9.9\%	&	2	&	-2.3	\\ 
& H7	&	5.0\%	&	2	&	-1.9	&	6.8\%	&	1	&	-2.9	\\ 
& H8	&	8.2\%	&	1	&	-11.5	&	54.7\%	&	4	&	-1.3	\\ 
& H9	&	2.9\%	&	1	&	-2.4	&	9.1\%	&	1	&	-1.3	\\ 
\hline
\multirow{5}{*}{\rotatebox[origin=c]{90}{\bf Office}} & O1	&	4.0\%	&	n.a.	&	n.a.	&	3.3\%	&	4	&	 0.4	\\ 
& O2	&	5.6\%	&	1	&	-7.9	&	2.9\%	&	2	&	-1.6	\\ 
& O3	&	1.7\%	&	1	&	-0.4	&	3.9\%	&	2	&	-0.3	\\ 
& O4	&	5.4\%	&	3	&	-2.5	&	5.8\%	&	2	&	-0.8	\\ 
& O5	&	4.8\%	&	6	&	-6.2	&	5.6\%	&	12	&	-0.2	\\ 
\hline
\multirow{5}{*}{\rotatebox[origin=c]{90}{\bf Corridor}} & C1	&	3.6\%	&	12	&	 0.1	&	4.4\%	&	11	&	 0.7	\\ 
& C2	&	6.2\%	&	2	&	 0.6	&	44.2\%	&	2	&	-1.0	\\ 
& C3	&	5.8\%	&	1	&	-7.6	&	4.4\%	&	1	&	-1.1	\\ 
& C4	&	1.9\%	&	1	&	-6.0	&	2.2\%	&	1	&	-2.8	\\ 
& C5	&	1.9\%	&	2	&	-3.7	&	5.8\%	&	1	&	-1.0	\\ 
\hline
	\end{tabular}

{\small * n.a. means that TouchSync cannot converge because of large $\epsilon$.}
\end{table}

\subsubsection{Laboratory} We conduct experiments in a computer science laboratory with about 100 seats and various office facilities (lights, computers, printers, projectors, meeting rooms, etc). Fig.~\ref{fig:cncl} shows the laboratory's floor plan. We arbitrarily select nine test points, marked by ``Lx'' in Fig.~\ref{fig:cncl}. A researcher carries the Floras to each test point and conducts two experiments. In the first experiment, the Floras have no physical contact with human body; in the second experiment,
the researcher wears the two Floras. Thus, the experiment evaluates the same-wearer scenario. The example applications mentioned in \sect\ref{sec:intro}, i.e., wireless earbuds, motion analysis, and muscle activation monitoring, belong to this scenario. Each synchronization session takes about $150\,\text{ms}$. A synchronization process completes once the IAS converges.

The first part of Table~\ref{tab:indoor-environments} shows the SEP's signal strength, the number of synchronization sessions until convergence ($K$), and the clock offset estimation error at each test point. Without skin contact, the signal strength is a few percent only. But TouchSync can still achieve a $3\,\text{ms}$ accuracy. At L7, TouchSync cannot converge because of large and varying $\epsilon$. The skin contact significantly increases the signal strength. Moreover, TouchSync converges after two synchronization sessions in most cases. For $K=2$, a synchronization process takes less than one second. The absolute clock offset estimation errors are below $1\,\text{ms}$, lower than those without skin contact. However, without skin contact, the accuracy does not substantially degrade. This suggests that, TouchSync is resilient to the loss of skin contact due to say loose wearing.

\begin{figure}
  \centering
  \includegraphics[width=.35\textwidth]{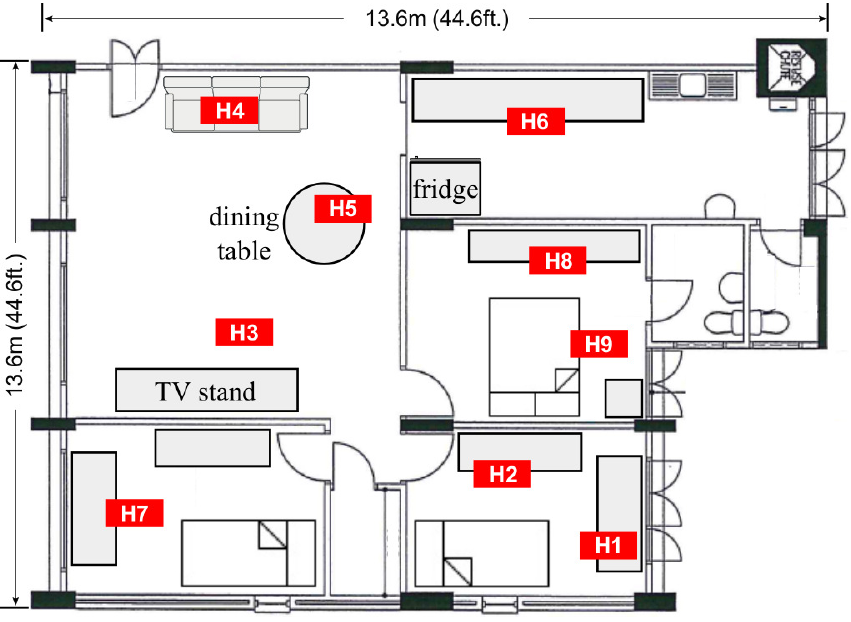}
  \caption{Home floor plan with test points marked.}
  \label{fig:home}
\end{figure}

\subsubsection{Home} We conduct experiments in a $104\,\text{m}^2$ three-bedroom home with typical home furniture and appliances. Fig.~\ref{fig:home} shows the home's floor plan. We arbitrarily select nine test points, marked by ``Hx'' in Fig.~\ref{fig:home}. The second part of Table~\ref{tab:indoor-environments} shows the results. Without skin contact, the signal strength results are similar to those obtained in the laboratory. With skin contact, the signal strength increases and the absolute clock offset estimation errors are below $3\,\text{ms}$.

\subsubsection{Office} The third part of Table~\ref{tab:indoor-environments} shows the results obtained at five test points in a $15\,\text{m}^2$ office. At test points O1 and O2, the signal strength with skin contact is slightly lower than that without skin contact. This is possible as the two tests were conducted during different time periods and the powerline radiation may vary over time due to changed electric currents. With skin contact, TouchSync gives sub-ms accuracy except at O2.

\subsubsection{Corridor} We select five test points with equal spacing in a $200\,\text{m}$ corridor of a campus building. The fourth part of Table~\ref{tab:indoor-environments} gives the results. With skin contact, TouchSync yields absolute clock offset estimation errors of about $1\,\text{ms}$ except at C4. The errors with skin contact are lower than those without skin contact.

\subsubsection{Discussions} In summary, with skin contact, TouchSync gives sub-ms clock offset estimation errors at 20 test points out of totally 28 test points in Table~\ref{tab:indoor-environments}. All errors are below $3\,\text{ms}$.

From Table~\ref{tab:indoor-environments}, at 3 out of 28 test points, the test without skin contact gives higher signal strength than that with skin contact. This is because the two tests are conducted sequentially and the EMR may change over time. Moreover, the signal strength exhibits significant variation at different locations. This is because the EMR decays with the distance from the powerline. Nevertheless,
the above results show the pervasive availability of SEP in indoor environments.
In our experiments presented above, we do not observe interference from electrical appliances that leads to wrong clock synchronization. In our future work, we will conduct extensive investigation on the impact of the possible electromagnetic radiation from various electrical appliances on TouchSync.

We now discuss the energy consumption of TouchSync. Compared with NTP, TouchSync's SEP sampling and extra synchronization sessions incur additional energy consumption. SEP sampling is performed during the clock synchronization process only. The ADC's power consumption is much lower than the radio's.
For instance, the ADC embedded in the TI MSP430G2x53 MCU consumes $0.6\,\text{mA}$ only, much lower than the power consumption of ZigBee and Bluetooth radios which is typically tens of mA.
Thus, the radio's energy consumption during the IAS's convergence process dominates TouchSync's energy consumption.
From Table 2, with skin contact, the IAS converges within 2.9 synchronization sessions on average. Thus, TouchSync's energy consumption is estimated as about three times of NTP's.

\subsection{TouchSync-over-Internet}

Tightly synchronizing wearables over long physical distances is often desirable. For instance, in distributed virtual reality applications, tight clock synchronization among participating sensing and rendering devices that may be geographically distributed is essential \cite{friedmann1992synchronization,hamza2004scene}.
Although the synchronization can be performed in a hop-by-hop manner (e.g., wearables $\leftrightarrow$ smartphone $\leftrightarrow$ cloud), errors accumulate over hops. In particular, tightly synchronizing a smartphone to global time has been a real and challenging problem -- tests showed that the synchronization through LTE and Wi-Fi experiences hundreds of ms jitters \cite{lazik2015ultrasonic}. In contrast, TouchSync can perform end-to-end synchronizations for wearables distributed in a geographic region served by the same power grid. The basis is that, the 50/60 Hz power grid voltage, which generates the powerline radiation, is highly synchronous across the whole power grid \cite{sreejaya16}. TouchSync can also synchronize wearables directly with a cloud server in the same region.
The smartphone merely relays the messages exchanged among the wearables and the cloud server if the wearables cannot directly access Internet.  The cloud server can use a sensor directly plugged in to a power outlet to capture the power grid voltage.
Owing to the Internet connectivity, the end-to-end synchronization scheme greatly simplifies the system design and implementation.

\begin{figure}
  \begin{minipage}[t]{.23\textwidth}
    \includegraphics{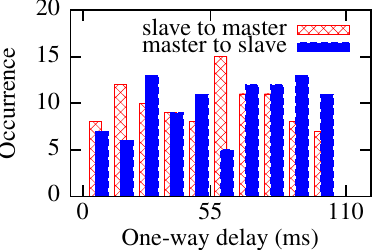}
    \caption{One-way delays over a ngrok tunnel.}
    \label{fig:ogrok-delay}
  \end{minipage}
  \hspace{0.2em}
  \begin{minipage}[t]{.23\textwidth}
    \includegraphics{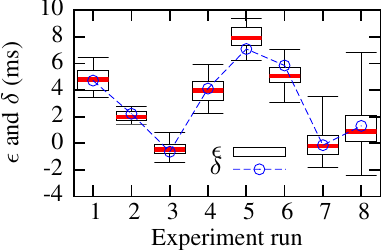}
    \caption{Accuracy of TouchSync-over-Internet.}
    \label{fig:touchsync-over-internet}
  \end{minipage}
\end{figure}

We conduct a proof-of-concept experiment of TouchSync-over-Internet as follows. Two researchers carry a Flora-RPi setup each to two buildings that are about $10\,\text{km}$ apart. The RPi is attached with an Adafruit GPS receiver to obtain ground-truth coordinated universal time (UTC) with $\mu$s accuracy. The two nodes, one as TouchSync master and the other as TouchSync slave, communicate through a tunnel established by ngrok 1.7, an open-source reverse proxy often adopted for creating IoT networks. Fig.~\ref{fig:ogrok-delay} shows the distributions of the two one-way delays over the ngrok tunnel. We can see that the ngrok exhibits greater dynamics than the BLE link (cf.~Fig.~\ref{fig:one-way-delays}). We evaluate TouchSync-over-Internet for eight times during a day. Fig.~\ref{fig:touchsync-over-internet} shows the box plots of the time displacements (i.e., $\epsilon$) between the SEPs captured by the two nodes. We can see that $\epsilon$ varies from $-2\,\text{ms}$ to $9\,\text{ms}$ during the day. From the building managements, the two rooms where the master and slave nodes are located draw electricity from the R and Y phases of the power grid, respectively. There is a phase difference of $20/3=6.7\,\text{ms}$ between these two phases. Moreover, from power engineering, the difference between the power grid voltage phases at different geographic locations is non-zero and time-varying. These factors lead to the non-zero and time-varying $\epsilon$ in Fig.~\ref{fig:touchsync-over-internet}. The dotted line in Fig.~\ref{fig:touchsync-over-internet} shows the synchronization errors of TouchSync-over-Internet (i.e., $\delta$) in various experiment runs. They are within the range of $\epsilon$, since $\epsilon$ is the major source of TouchSync's synchronization error. The largest $\delta$ is $7\,\text{ms}$. The integer ambiguity solver converges within 4 to 13 synchronization sessions.

\section{Conclusion}
\label{sec:conclude}

TouchSync synchronizes the clocks of wearables by exploiting the wearers' skin electric potentials induced by powerline radiation. Different from existing WSN clock synchronization approaches that find difficulties in being applied on diverse IoT platforms due to their need of hardware-level packet timestamping or non-trivial extra hardware, TouchSync can be readily implemented as an app based on standard wearable OS calls. Extensive evaluation shows TouchSync's synchronization errors of below $3\,\text{ms}$ and $7\,\text{ms}$ on the same wearer and between two wearers $10\,\text{km}$ apart, respectively.

\begin{acks}
The authors wish to thank our shepherd Dr. Lu Su and the anonymous reviewers for providing valuable feedback on this work. This research was funded by a Start-up Grant at Nanyang Technological University.
\end{acks}
\balance

\bibliographystyle{ACM-Reference-Format}

\bibliography{reference} 


\begin{thebibliography}{00}


\ifx \showCODEN    \undefined \def \showCODEN     #1{\unskip}     \fi
\ifx \showDOI      \undefined \def \showDOI       #1{#1}\fi
\ifx \showISBNx    \undefined \def \showISBNx     #1{\unskip}     \fi
\ifx \showISBNxiii \undefined \def \showISBNxiii  #1{\unskip}     \fi
\ifx \showISSN     \undefined \def \showISSN      #1{\unskip}     \fi
\ifx \showLCCN     \undefined \def \showLCCN      #1{\unskip}     \fi
\ifx \shownote     \undefined \def \shownote      #1{#1}          \fi
\ifx \showarticletitle \undefined \def \showarticletitle #1{#1}   \fi
\ifx \showURL      \undefined \def \showURL       {\relax}        \fi
\providecommand\bibfield[2]{#2}
\providecommand\bibinfo[2]{#2}
\providecommand\natexlab[1]{#1}
\providecommand\showeprint[2][]{arXiv:#2}

\bibitem[\protect\citeauthoryear{Chan, Est\`{e}Ve, Fourniols, Escriba, and
  Campo}{Chan et~al\mbox{.}}{2012}]%
        {chan2012smart}
\bibfield{author}{\bibinfo{person}{Marie Chan}, \bibinfo{person}{Daniel
  Est\`{e}Ve}, \bibinfo{person}{Jean-Yves Fourniols},
  \bibinfo{person}{Christophe Escriba}, {and} \bibinfo{person}{Eric Campo}.}
  \bibinfo{year}{2012}\natexlab{}.
\newblock \showarticletitle{Smart Wearable Systems: Current Status and Future
  Challenges}.
\newblock \bibinfo{journal}{{\em Artificial Intelligence in Medicine\/}}
  \bibinfo{volume}{56}, \bibinfo{number}{3} (\bibinfo{date}{Nov.}
  \bibinfo{year}{2012}), \bibinfo{pages}{137--156}.
\newblock
\showISSN{0933-3657}


\bibitem[\protect\citeauthoryear{Chen, Wang, Chang, and Terzis}{Chen
  et~al\mbox{.}}{2011}]%
        {Chen2011ultralow}
\bibfield{author}{\bibinfo{person}{Yin Chen}, \bibinfo{person}{Qiang Wang},
  \bibinfo{person}{Marcus Chang}, {and} \bibinfo{person}{Andreas Terzis}.}
  \bibinfo{year}{2011}\natexlab{}.
\newblock \showarticletitle{Ultra-low power time synchronization using passive
  radio receivers}. In \bibinfo{booktitle}{{\em Proceedings of the 10th
  ACM/IEEE International Conference on Information Processing in Sensor
  Networks (IPSN)}}. \bibinfo{publisher}{IEEE}, \bibinfo{address}{Chicago, IL,
  USA}, \bibinfo{pages}{235--245}.
\newblock


\bibitem[\protect\citeauthoryear{Dinescu, Mazza, Kujanski, Gaza, and
  Sagan}{Dinescu et~al\mbox{.}}{2015}]%
        {dinescu2015synchronizing}
\bibfield{author}{\bibinfo{person}{Mihail~C. Dinescu}, \bibinfo{person}{Joseph
  Mazza}, \bibinfo{person}{Adam Kujanski}, \bibinfo{person}{Brian Gaza}, {and}
  \bibinfo{person}{Michael Sagan}.} \bibinfo{year}{2015}\natexlab{}.
\newblock \bibinfo{title}{Synchronizing wireless earphones}.
\newblock   (\bibinfo{date}{April~7} \bibinfo{year}{2015}).
\newblock
\showURL{%
\url{https://www.google.com/patents/US9002044}}
\newblock
\shownote{US Patent 9,002,044.}


\bibitem[\protect\citeauthoryear{Elson, Girod, and Estrin}{Elson
  et~al\mbox{.}}{2002}]%
        {elson2002fine}
\bibfield{author}{\bibinfo{person}{Jeremy Elson}, \bibinfo{person}{Lewis
  Girod}, {and} \bibinfo{person}{Deborah Estrin}.}
  \bibinfo{year}{2002}\natexlab{}.
\newblock \showarticletitle{Fine-grained Network Time Synchronization Using
  Reference Broadcasts}.
\newblock \bibinfo{journal}{{\em SIGOPS Operating Systems Review\/}}
  \bibinfo{volume}{36}, \bibinfo{number}{SI} (\bibinfo{date}{Dec.}
  \bibinfo{year}{2002}), \bibinfo{pages}{147--163}.
\newblock
\showISSN{0163-5980}


\bibitem[\protect\citeauthoryear{Foundation}{Foundation}{2017}]%
        {rpi3}
\bibfield{author}{\bibinfo{person}{Raspberry~Pi Foundation}.}
  \bibinfo{year}{2017}\natexlab{}.
\newblock \bibinfo{title}{Raspberry {Pi} 3 {Model} B}.
\newblock   (\bibinfo{year}{2017}).
\newblock
\newblock
\shownote{\url{https://www.raspberrypi.org/products/raspberry-pi-3-model-b/}.}


\bibitem[\protect\citeauthoryear{Friedmann, Starner, and Pentland}{Friedmann
  et~al\mbox{.}}{1992}]%
        {friedmann1992synchronization}
\bibfield{author}{\bibinfo{person}{Martin Friedmann}, \bibinfo{person}{Thad
  Starner}, {and} \bibinfo{person}{Alex Pentland}.}
  \bibinfo{year}{1992}\natexlab{}.
\newblock \showarticletitle{Synchronization in Virtual Realities}.
\newblock \bibinfo{journal}{{\em Presence: Teleoperators and Virtual
  Environments\/}} \bibinfo{volume}{1}, \bibinfo{number}{1}
  (\bibinfo{date}{Jan.} \bibinfo{year}{1992}), \bibinfo{pages}{139--144}.
\newblock
\showISSN{1054-7460}


\bibitem[\protect\citeauthoryear{Ganeriwal, Kumar, and Srivastava}{Ganeriwal
  et~al\mbox{.}}{2003}]%
        {ganeriwal2003timing}
\bibfield{author}{\bibinfo{person}{Saurabh Ganeriwal}, \bibinfo{person}{Ram
  Kumar}, {and} \bibinfo{person}{Mani~B. Srivastava}.}
  \bibinfo{year}{2003}\natexlab{}.
\newblock \showarticletitle{Timing-sync Protocol for Sensor Networks}. In
  \bibinfo{booktitle}{{\em Proceedings of the 1st International Conference on
  Embedded Networked Sensor Systems}} {\em (\bibinfo{series}{SenSys})}.
  \bibinfo{publisher}{ACM}, \bibinfo{address}{Los Angeles, California, USA},
  \bibinfo{pages}{138--149}.
\newblock
\showISBNx{1-58113-707-9}


\bibitem[\protect\citeauthoryear{Geboff, Hariharan, Linde, Tan, and
  Shahparnia}{Geboff et~al\mbox{.}}{2015}]%
        {apple-sync}
\bibfield{author}{\bibinfo{person}{Adam Geboff}, \bibinfo{person}{Sriram
  Hariharan}, \bibinfo{person}{Joakim Linde}, \bibinfo{person}{Li-Quan Tan},
  {and} \bibinfo{person}{Shahrooz Shahparnia}.}
  \bibinfo{year}{2015}\natexlab{}.
\newblock \bibinfo{title}{Device synchronization over bluetooth}.
\newblock   (\bibinfo{date}{April~2} \bibinfo{year}{2015}).
\newblock
\showURL{%
\url{https://www.google.com/patents/US20150092642}}
\newblock
\shownote{US Patent App. 14/496,314.}


\bibitem[\protect\citeauthoryear{Hamza-Lup and Rolland}{Hamza-Lup and
  Rolland}{2004}]%
        {hamza2004scene}
\bibfield{author}{\bibinfo{person}{Felix~G. Hamza-Lup} {and}
  \bibinfo{person}{Jannick~P. Rolland}.} \bibinfo{year}{2004}\natexlab{}.
\newblock \showarticletitle{Scene Synchronization for Real-time Interaction in
  Distributed Mixed Reality and Virtual Reality Environments}.
\newblock \bibinfo{journal}{{\em Presence: Teleoperators and Virtual
  Environments\/}} \bibinfo{volume}{13}, \bibinfo{number}{3}
  (\bibinfo{date}{July} \bibinfo{year}{2004}), \bibinfo{pages}{315--327}.
\newblock
\showISSN{1054-7460}


\bibitem[\protect\citeauthoryear{Hao, Zhou, Xing, and Mutka}{Hao
  et~al\mbox{.}}{2011}]%
        {hao2011wizsync}
\bibfield{author}{\bibinfo{person}{Tian Hao}, \bibinfo{person}{Ruogu Zhou},
  \bibinfo{person}{Guoliang Xing}, {and} \bibinfo{person}{Matt Mutka}.}
  \bibinfo{year}{2011}\natexlab{}.
\newblock \showarticletitle{WizSync: Exploiting Wi-Fi Infrastructure for Clock
  Synchronization in Wireless Sensor Networks}. In \bibinfo{booktitle}{{\em
  Proceedings of the 32nd Real-Time Systems Symposium (RTSS)}}.
  \bibinfo{publisher}{IEEE}, \bibinfo{address}{Vienna, Austria},
  \bibinfo{pages}{149--158}.
\newblock
\showISSN{1052-8725}


\bibitem[\protect\citeauthoryear{IDC}{IDC}{2016}]%
        {idc-wearable}
\bibfield{author}{\bibinfo{person}{IDC}.} \bibinfo{year}{2016}\natexlab{}.
\newblock \bibinfo{title}{{IDC} Forecasts Wearables Shipments to Reach 213.6
  Million Units Worldwide in 2020}.
\newblock   (\bibinfo{year}{2016}).
\newblock
\newblock
\shownote{\url{http://www.idc.com/getdoc.jsp?containerId=prUS41530816}.}


\bibitem[\protect\citeauthoryear{IEEE}{IEEE}{2008}]%
        {4579760}
\bibfield{author}{\bibinfo{person}{IEEE}.} \bibinfo{year}{2008}\natexlab{}.
\newblock \showarticletitle{IEEE Standard for a Precision Clock Synchronization
  Protocol for Networked Measurement and Control Systems}.
\newblock \bibinfo{journal}{{\em IEEE Std 1588-2008 (Revision of IEEE Std
  1588-2002)\/}} \bibinfo{volume}{1}, \bibinfo{number}{1} (\bibinfo{date}{July}
  \bibinfo{year}{2008}), \bibinfo{pages}{1--300}.
\newblock
\showDOI{%
\url{https://doi.org/10.1109/IEEESTD.2008.4579760}}


\bibitem[\protect\citeauthoryear{Inc}{Inc}{2017}]%
        {z1}
\bibfield{author}{\bibinfo{person}{Zolertia Inc}.}
  \bibinfo{year}{2017}\natexlab{}.
\newblock \bibinfo{title}{Z1 mote}.
\newblock   (\bibinfo{year}{2017}).
\newblock
\newblock
\shownote{\url{http://zolertia.io/z1}.}


\bibitem[\protect\citeauthoryear{Industries}{Industries}{2017a}]%
        {flora}
\bibfield{author}{\bibinfo{person}{Adafruit Industries}.}
  \bibinfo{year}{2017}\natexlab{a}.
\newblock \bibinfo{title}{Adafruit {FLORA}}.
\newblock   (\bibinfo{year}{2017}).
\newblock
\newblock
\shownote{\url{https://www.adafruit.com/category/92}.}


\bibitem[\protect\citeauthoryear{Industries}{Industries}{2017b}]%
        {adafruit-nrf51}
\bibfield{author}{\bibinfo{person}{Adafruit Industries}.}
  \bibinfo{year}{2017}\natexlab{b}.
\newblock \bibinfo{title}{Adafruit {nRF51} {BLE} Library}.
\newblock   (\bibinfo{year}{2017}).
\newblock
\newblock
\shownote{\url{https://learn.adafruit.com/adafruit-feather-32u4-bluefruit-le/installing-ble-library}.}


\bibitem[\protect\citeauthoryear{Industries}{Industries}{2017c}]%
        {adafruit-bluefruitle}
\bibfield{author}{\bibinfo{person}{Adafruit Industries}.}
  \bibinfo{year}{2017}\natexlab{c}.
\newblock \bibinfo{title}{Adafruit {Python} {BluefruitLE}}.
\newblock   (\bibinfo{year}{2017}).
\newblock
\newblock
\shownote{\url{https://github.com/adafruit/Adafruit_Python_BluefruitLE}.}


\bibitem[\protect\citeauthoryear{Klepeis, Nelson, Ott, Robinson, Tsang,
  Switzer, Behar, Hern, and Engelmann}{Klepeis et~al\mbox{.}}{2001}]%
        {klepeis2001national}
\bibfield{author}{\bibinfo{person}{Neil~E. Klepeis},
  \bibinfo{person}{William~C. Nelson}, \bibinfo{person}{Wayne~R. Ott},
  \bibinfo{person}{John~P. Robinson}, \bibinfo{person}{Andy~M. Tsang},
  \bibinfo{person}{Paul Switzer}, \bibinfo{person}{Joseph~V. Behar},
  \bibinfo{person}{Stephen~C. Hern}, {and} \bibinfo{person}{William~H.
  Engelmann}.} \bibinfo{year}{2001}\natexlab{}.
\newblock \showarticletitle{The National Human Activity Pattern Survey (NHAPS):
  a resource for assessing exposure to environmental pollutants}.
\newblock \bibinfo{journal}{{\em Journal of Exposure Analysis and Environmental
  Epidemiology\/}} \bibinfo{volume}{11}, \bibinfo{number}{3}
  (\bibinfo{year}{2001}), \bibinfo{pages}{231--252}.
\newblock
\showISSN{1053-4245}


\bibitem[\protect\citeauthoryear{Lazik, Rajagopal, Sinopoli, and Rowe}{Lazik
  et~al\mbox{.}}{2015}]%
        {lazik2015ultrasonic}
\bibfield{author}{\bibinfo{person}{Patrick Lazik}, \bibinfo{person}{Niranjini
  Rajagopal}, \bibinfo{person}{Bruno Sinopoli}, {and} \bibinfo{person}{Anthony
  Rowe}.} \bibinfo{year}{2015}\natexlab{}.
\newblock \showarticletitle{Ultrasonic time synchronization and ranging on
  smartphones}. In \bibinfo{booktitle}{{\em Proceedings of 21st IEEE Real-Time
  and Embedded Technology and Applications Symposium (RTAS)}}.
  \bibinfo{publisher}{IEEE}, \bibinfo{address}{Seattle, WA, USA},
  \bibinfo{pages}{108--118}.
\newblock
\showISSN{1545-3421}


\bibitem[\protect\citeauthoryear{Li, Xing, Sun, Huangfu, Zhou, and Zhu}{Li
  et~al\mbox{.}}{2011}]%
        {li2011exploiting}
\bibfield{author}{\bibinfo{person}{Liqun Li}, \bibinfo{person}{Guoliang Xing},
  \bibinfo{person}{Limin Sun}, \bibinfo{person}{Wei Huangfu},
  \bibinfo{person}{Ruogu Zhou}, {and} \bibinfo{person}{Hongsong Zhu}.}
  \bibinfo{year}{2011}\natexlab{}.
\newblock \showarticletitle{Exploiting FM Radio Data System for Adaptive Clock
  Calibration in Sensor Networks}. In \bibinfo{booktitle}{{\em Proceedings of
  the 9th International Conference on Mobile Systems, Applications, and
  Services}} {\em (\bibinfo{series}{MobiSys})}. \bibinfo{publisher}{ACM},
  \bibinfo{address}{Bethesda, Maryland, USA}, \bibinfo{pages}{169--182}.
\newblock
\showISBNx{978-1-4503-0643-0}


\bibitem[\protect\citeauthoryear{Li, Tan, and Yau}{Li et~al\mbox{.}}{2017}]%
        {yang17ipsn}
\bibfield{author}{\bibinfo{person}{Yang Li}, \bibinfo{person}{Rui Tan}, {and}
  \bibinfo{person}{David K.~Y. Yau}.} \bibinfo{year}{2017}\natexlab{}.
\newblock \showarticletitle{Natural Timestamping Using Powerline
  Electromagnetic Radiation}. In \bibinfo{booktitle}{{\em Proceedings of the
  16th ACM/IEEE International Conference on Information Processing in Sensor
  Networks}} {\em (\bibinfo{series}{IPSN})}. \bibinfo{publisher}{ACM},
  \bibinfo{address}{Pittsburgh, Pennsylvania}, \bibinfo{pages}{55--66}.
\newblock
\showISBNx{978-1-4503-4890-4}


\bibitem[\protect\citeauthoryear{Li, Chen, Li, Li, Li, and Liu}{Li
  et~al\mbox{.}}{2012}]%
        {li2012flight}
\bibfield{author}{\bibinfo{person}{Zhenjiang Li}, \bibinfo{person}{Wenwei
  Chen}, \bibinfo{person}{Cheng Li}, \bibinfo{person}{Mo Li},
  \bibinfo{person}{Xiang-Yang Li}, {and} \bibinfo{person}{Yunhao Liu}.}
  \bibinfo{year}{2012}\natexlab{}.
\newblock \showarticletitle{FLIGHT: Clock Calibration Using Fluorescent
  Lighting}. In \bibinfo{booktitle}{{\em Proceedings of the 18th Annual
  International Conference on Mobile Computing and Networking}} {\em
  (\bibinfo{series}{MobiCom})}. \bibinfo{publisher}{ACM},
  \bibinfo{address}{Istanbul, Turkey}, \bibinfo{pages}{329--340}.
\newblock
\showISBNx{978-1-4503-1159-5}


\bibitem[\protect\citeauthoryear{Lorincz, Chen, Challen, Chowdhury, Patel,
  Bonato, and Welsh}{Lorincz et~al\mbox{.}}{2009}]%
        {lorincz2009mercury}
\bibfield{author}{\bibinfo{person}{Konrad Lorincz}, \bibinfo{person}{Bor-rong
  Chen}, \bibinfo{person}{Geoffrey~Werner Challen}, \bibinfo{person}{Atanu~Roy
  Chowdhury}, \bibinfo{person}{Shyamal Patel}, \bibinfo{person}{Paolo Bonato},
  {and} \bibinfo{person}{Matt Welsh}.} \bibinfo{year}{2009}\natexlab{}.
\newblock \showarticletitle{Mercury: A Wearable Sensor Network Platform for
  High-fidelity Motion Analysis}. In \bibinfo{booktitle}{{\em Proceedings of
  the 7th ACM Conference on Embedded Networked Sensor Systems}} {\em
  (\bibinfo{series}{SenSys})}. \bibinfo{publisher}{ACM},
  \bibinfo{address}{Berkeley, California}, \bibinfo{pages}{183--196}.
\newblock
\showISBNx{978-1-60558-519-2}


\bibitem[\protect\citeauthoryear{Mar\'{o}ti, Kusy, Simon, and
  L{\'e}deczi}{Mar\'{o}ti et~al\mbox{.}}{2004}]%
        {maroti2004flooding}
\bibfield{author}{\bibinfo{person}{Mikl\'{o}s Mar\'{o}ti},
  \bibinfo{person}{Branislav Kusy}, \bibinfo{person}{Gyula Simon}, {and}
  \bibinfo{person}{\'{A}kos L{\'e}deczi}.} \bibinfo{year}{2004}\natexlab{}.
\newblock \showarticletitle{The Flooding Time Synchronization Protocol}. In
  \bibinfo{booktitle}{{\em Proceedings of the 2nd International Conference on
  Embedded Networked Sensor Systems}} {\em (\bibinfo{series}{SenSys})}.
  \bibinfo{publisher}{ACM}, \bibinfo{address}{Baltimore, MD, USA},
  \bibinfo{pages}{39--49}.
\newblock
\showISBNx{1-58113-879-2}


\bibitem[\protect\citeauthoryear{Mokaya, Lucas, Noh, and Zhang}{Mokaya
  et~al\mbox{.}}{2015}]%
        {mokaya2015myovibe}
\bibfield{author}{\bibinfo{person}{Frank Mokaya}, \bibinfo{person}{Roland
  Lucas}, \bibinfo{person}{Hae~Young Noh}, {and} \bibinfo{person}{Pei Zhang}.}
  \bibinfo{year}{2015}\natexlab{}.
\newblock \showarticletitle{MyoVibe: Vibration Based Wearable Muscle Activation
  Detection in High Mobility Exercises}. In \bibinfo{booktitle}{{\em
  Proceedings of the 2015 ACM International Joint Conference on Pervasive and
  Ubiquitous Computing}} {\em (\bibinfo{series}{UbiComp})}.
  \bibinfo{publisher}{ACM}, \bibinfo{address}{Osaka, Japan},
  \bibinfo{pages}{27--38}.
\newblock
\showISBNx{978-1-4503-3574-4}


\bibitem[\protect\citeauthoryear{Mokaya, Lucas, Noh, and Zhang}{Mokaya
  et~al\mbox{.}}{2016}]%
        {mokaya2016burnout}
\bibfield{author}{\bibinfo{person}{Frank Mokaya}, \bibinfo{person}{Roland
  Lucas}, \bibinfo{person}{Hae~Young Noh}, {and} \bibinfo{person}{Pei Zhang}.}
  \bibinfo{year}{2016}\natexlab{}.
\newblock \showarticletitle{Burnout: A Wearable System for Unobtrusive Skeletal
  Muscle Fatigue Estimation}. In \bibinfo{booktitle}{{\em Proceedings of the
  15th International Conference on Information Processing in Sensor Networks}}
  {\em (\bibinfo{series}{IPSN})}. \bibinfo{publisher}{IEEE},
  \bibinfo{address}{Vienna, Austria}, \bibinfo{pages}{1--12}.
\newblock
\showISBNx{978-1-5090-0802-5}


\bibitem[\protect\citeauthoryear{Project}{Project}{2017}]%
        {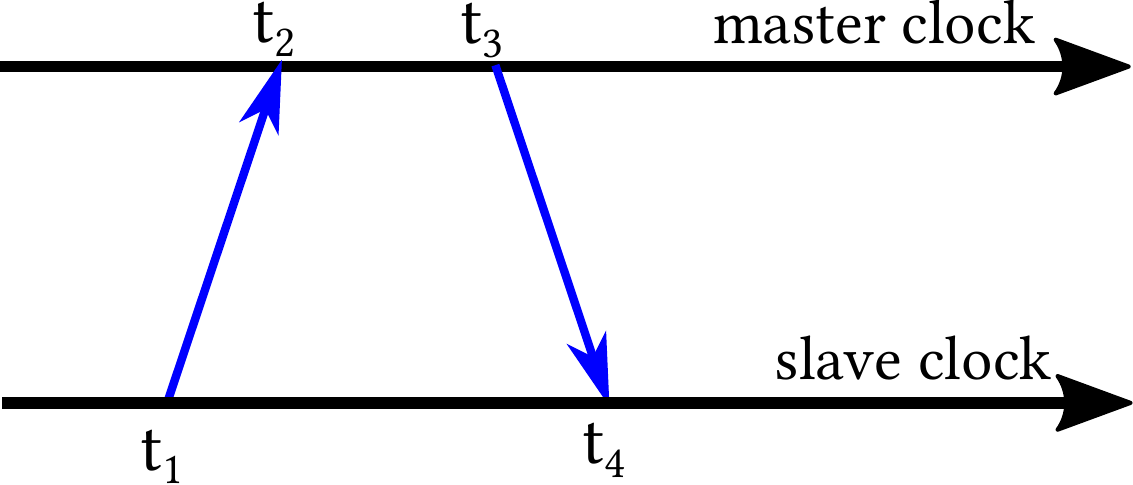}
\bibfield{author}{\bibinfo{person}{NTP Project}.}
  \bibinfo{year}{2017}\natexlab{}.
\newblock \bibinfo{title}{{NTP}: The Network Time Protocol}.
\newblock   (\bibinfo{year}{2017}).
\newblock
\newblock
\shownote{\url{http://www.ntp.org/}.}


\bibitem[\protect\citeauthoryear{Rabadi, Tan, Yau, and Viswanathan}{Rabadi
  et~al\mbox{.}}{2017}]%
        {dima17}
\bibfield{author}{\bibinfo{person}{Dima Rabadi}, \bibinfo{person}{Rui Tan},
  \bibinfo{person}{David~K.Y. Yau}, {and} \bibinfo{person}{Sreejaya
  Viswanathan}.} \bibinfo{year}{2017}\natexlab{}.
\newblock \showarticletitle{Taming Asymmetric Network Delays for Clock
  Synchronization Using Power Grid Voltage}. In \bibinfo{booktitle}{{\em
  Proceedings of the 2017 ACM on Asia Conference on Computer and Communications
  Security}} {\em (\bibinfo{series}{ASIACCS})}. \bibinfo{publisher}{ACM},
  \bibinfo{address}{Abu Dhabi, United Arab Emirates},
  \bibinfo{pages}{874--886}.
\newblock
\showISBNx{978-1-4503-4944-4}


\bibitem[\protect\citeauthoryear{Rowe, Gupta, and Rajkumar}{Rowe
  et~al\mbox{.}}{2009}]%
        {Rowe2009lowpower}
\bibfield{author}{\bibinfo{person}{Anthony Rowe}, \bibinfo{person}{Vikram
  Gupta}, {and} \bibinfo{person}{Ragunathan~(Raj) Rajkumar}.}
  \bibinfo{year}{2009}\natexlab{}.
\newblock \showarticletitle{Low-power Clock Synchronization Using
  Electromagnetic Energy Radiating from AC Power Lines}. In
  \bibinfo{booktitle}{{\em Proceedings of the 7th ACM Conference on Embedded
  Networked Sensor Systems}} {\em (\bibinfo{series}{SenSys})}.
  \bibinfo{publisher}{ACM}, \bibinfo{address}{Berkeley, California},
  \bibinfo{pages}{211--224}.
\newblock
\showISBNx{978-1-60558-519-2}


\bibitem[\protect\citeauthoryear{{Symmetricom, Inc.}}{{Symmetricom,
  Inc.}}{2011}]%
        {chipatomic}
\bibfield{author}{\bibinfo{person}{{Symmetricom, Inc.}}}
  \bibinfo{year}{2011}\natexlab{}.
\newblock \bibinfo{title}{Symmetricom annouces general availability of
  industry's first commercially-available chip scale atomtic clock}.
\newblock   (\bibinfo{year}{2011}).
\newblock
\newblock
\shownote{\url{http://bit.ly/2yzQ32P}.}


\bibitem[\protect\citeauthoryear{Viswanathan, Tan, and Yau}{Viswanathan
  et~al\mbox{.}}{2016}]%
        {sreejaya16}
\bibfield{author}{\bibinfo{person}{Sreejaya Viswanathan}, \bibinfo{person}{Rui
  Tan}, {and} \bibinfo{person}{David~K.Y. Yau}.}
  \bibinfo{year}{2016}\natexlab{}.
\newblock \showarticletitle{Exploiting Power Grid for Accurate and Secure Clock
  Synchronization in Industrial IoT}. In \bibinfo{booktitle}{{\em Proceedings
  of the 37th IEEE Real-Time Systems Symposium (RTSS)}}.
  \bibinfo{publisher}{IEEE}, \bibinfo{address}{Porto, Portugal},
  \bibinfo{pages}{146--156}.
\newblock


\bibitem[\protect\citeauthoryear{Yan, Li, Tan, and Jun}{Yan
  et~al\mbox{.}}{2017}]%
        {touchsync-h}
\bibfield{author}{\bibinfo{person}{Zhenyu Yan}, \bibinfo{person}{Yang Li},
  \bibinfo{person}{Rui Tan}, {and} \bibinfo{person}{Huang Jun}.}
  \bibinfo{year}{2017}\natexlab{}.
\newblock \bibinfo{title}{TouchSync implementation}.
\newblock   (\bibinfo{year}{2017}).
\newblock
\newblock
\shownote{\url{https://github.com/yanmarvin/touchsync}.}


\end{thebibliography}

\end{document}